\documentclass[prd,twocolumn,showpacs, floatfix, nofootinbib,
preprintnumbers
]{revtex4}
\usepackage{amsmath}
\usepackage{amssymb}
\usepackage{epsfig}
\usepackage{graphicx}
\usepackage{color}
\usepackage{hyperref}
\usepackage{dsfont}

\newcommand{\be}{\begin{equation}}
\newcommand{\ee}{\end{equation}}
\newcommand{\bdm}{\begin{displaymath}}
\newcommand{\edm}{\end{displaymath}}
\newcommand{\bea}{\begin{eqnarray}}
\newcommand{\eea}{\end{eqnarray}}
\newcommand{\nn}{\nonumber}

\newcommand{\vev}[1]{\langle #1 \rangle}
\newcommand{\unit}{\mathds{1}}
\newcommand{\PMNS}{V_{PMNS}}
\newcommand{\PMNSx}{V}

\begin{document}
\title{Witten's mechanism in the flipped $SU(5)$ unification}
\preprint{IFIC/13-65}
\author{Carolina Arbel\'aez}\email{carolina.arbelaez@ist.utl.pt}
\affiliation{Centro de F\'isica Te\'orica de Part\'iculas, Instituto Superior T\'ecnico, Universidade T\'ecnica de Lisboa, Av. Rovisco Pais 1, 1049-001, Lisboa Portugal \\ and \\AHEP Group, Instituto de F\'\i sica Corpuscular -- C.S.I.C./Universitat de Val\`encia Edificio de Institutos de Paterna, Apartado 22085, E--46071 Val\`encia, Spain}
\author{Helena Kole\v{s}ov\'a}\email{helena.sedivakova@fjfi.cvut.cz}
\affiliation{Institute of Particle and Nuclear Physics,
Faculty of Mathematics and Physics,
Charles University in Prague, V Hole\v{s}ovi\v{c}k\'ach 2,
180 00 Praha 8, Czech Republic \\ and \\Faculty of Nuclear Sciences and Physical Engineering, Czech Technical University in Prague, B\v{r}ehov\'a 7, 115 19 Praha 1, Czech Republic
}
\author{Michal Malinsk\'{y}}\email{malinsky@ipnp.troja.mff.cuni.cz}
\affiliation{Institute of Particle and Nuclear Physics,
Faculty of Mathematics and Physics,
Charles University in Prague, V Hole\v{s}ovi\v{c}k\'ach 2,
180 00 Praha 8, Czech Republic}
\begin{abstract}
We argue that Witten's loop mechanism for the right-handed Majorana neutrino mass generation  identified originally in the $SO(10)$ grand unification context can be successfully adopted to the class of the simplest flipped $SU(5)$ models. In such a framework, the main drawback of the $SO(10)$ prototype\nobreakdash ---in particular, the generic tension among the gauge unification constraints and the absolute neutrino mass scale\nobreakdash ---is alleviated and a simple yet potentially realistic and testable scenario emerges.
\end{abstract}
\maketitle
\section{Introduction}
The apparent absence of supersymmetry in the sub-TeV domain indicated by the current LHC data reopens the question whether the unprecedented smallness of the absolute neutrino mass scale may be ascribed to a loop suppression with the underlying dynamics in the TeV ballpark rather than the traditional seesaw~\cite{Minkowski:1977sc,Yanagida:1979as,Mohapatra:1979ia,Schechter:1980gr,Lazarides:1980nt,Foot:1988aq} picture featuring a very high scale, typically far beyond our reach. Recently, there has been a lot of activity in this direction with, e.g., dedicated studies of the Zee \cite{Zee:1980ai}, Zee-Babu \cite{Zee:1985rj,Zee:1985id,Babu:1988ki} and other models (cf. \cite{Bonnet:2012kz,Angel:2012ug} and references therein) focusing on their distinctive low-energy phenomenology and, in particular, their potential to be probed at the LHC and other facilities, see, e.g.,~\cite{Baek:2012ub,Ohlsson:2009vk,Nebot:2007bc,AristizabalSierra:2006gb,Frampton:2001eu}.

With the upcoming generation of megaton-scale experiments \cite{Abe:2011ts,Akiri:2011dv,Autiero:2007zj} dedicated, besides precision neutrino physics, to the search of perturbative baryon number violating (BNV) processes such as proton decay, the same question can be readdressed from the high-energy perspective. In principle, there can be high-scale loop diagrams behind the right-handed (RH) neutrino masses underpinning the seesaw mechanism rather than a direct low-scale $LL$ contraction, with possible imprints in the BNV physics.

Among such options, a prominent role is played by  Witten's scheme~\cite{Witten:1979nr} in the framework of the $SO(10)$ grand unification (GUT) where a pair of lepton-number violating vacuum expectation values (VEVs) is tied to the leptonic sector at two loops. Its main beauty consists in the observation that the RH neutrino masses are generated at the renormalizable level even in the simplest realization of $SO(10)$ with just the minimal scalar contents sufficient for the desired spontaneous symmetry breaking (i.e., $10\oplus 16\oplus 45$, cf.~\cite{Bertolini:2009es} and references therein); hence, there is in principle no need to invoke large scalar representations for that sake.

In practice, however,  Witten's mechanism has never found a clearly natural realization as a basis for a potentially realistic model building.
Among the possible reasons there is, namely, the dichotomy between the gauge unification constraints and the absolute size of Witten's loop governed by the position of the $B-L$ breaking scale $M_{B-L}$ which is required to be around the GUT-scale ($M_{G}$), due to the $(\alpha/\pi)^{2}$ suppression factor, in order to yield the ``correct'' seesaw scale $M_{R}\sim (\alpha/\pi)^{2} M_{B-L}^{2}/M_{G}$ in the $10^{13}$ GeV ballpark.
On one hand, this is exactly the situation encountered in supersymmetric GUTs where the one-step breaking picture characterized by a close proximity of $M_{B-L}$ and $M_{G}$ is essentially inevitable; at the same time, however, the low-scale supersymmetry makes the $F$-type loops at the GUT scale entirely academic due to the large cancellation involved.
On the other hand, non-SUSY GUTs generally require $M_{B-L}\ll M_{G}$ in order to account for the gauge unification constraints for which  Witten's mechanism yields contribution much below the desired $M_{R}\sim 10^{13}$ GeV.

In this respect, the beginning of the 1980s, when the low-energy SUSY was not yet mainstream and the lack of  detailed information about the standard model (SM) gauge coupling evolution as well as the absolute light neutrino mass scale obscured the issue with the too low Witten's $M_{R}$ in non-SUSY  scenarios, was the only time when this business really flourished\footnote{This can be seen at the citation counts of the original study~\cite{Witten:1979nr} as about 70\% of its today's total dates back to before 1985.}.  For a more recent attempt to implement such ideas in a simple, yet potentially realistic scenario the reader is  deferred to, e.g., the works \cite{Bajc:2004hr,Bajc:2005aq} where the split supersymmetry scheme supports both $M_{B-L}\sim M_{G}$ and very heavy scalar superpartners for which, in turn, the GUT-scale $F$-type Witten's loop is not entirely canceled.

In this work we approach this conundrum from a different perspective; in particular, we stick to the core of Witten's loop while relaxing, at the same time, the strict gauge unification constraints. For that sake, we depart from the canonical realization of Witten's mechanism in a full-fledged $SO(10)$ GUT to its ``bare-bone'' version which, as we point out, can be sensibly implemented within its simpler cousin, namely, the flipped $SU(5)$~\cite{Derendinger:1983aj,PhysRevLett.45.413,Barr:1981qv}.

Indeed, the strict full gauge unification constraints inherent to the $SO(10)$ GUTs are relaxed in such a scenario [owing to the nonsimple structure of its $SU(5)\otimes U(1)$ gauge group] which, in turn, makes it possible to have the rank-reducing vacuum expectation value (VEV) governing Witten's loop in the $10^{16}$~GeV ballpark even if the theory is  nonsupersymmetric.

The reason we are focusing just on the flipped $SU(5)$ framework is twofold: First, the baryon-number violating observables such as the $d=6$ proton decay~\cite{Nath:2006ut} may still be used to constrain specific scenarios even if the underlying dynamics is as high as at $10^{16}$~GeV, as we will comment upon in the following. This virtue is obviously lost if one picks any of the ``smaller'' subgroups of $SO(10)$ such as Pati-Salam\footnote{Let us recall that proton decay in Pati-Salam requires a conspiracy in the Higgs sector as it does not run solely through the gauge interactions.}~\cite{Pati:1974yy}, let alone the number of left-right symmetric (LR) settings based on the $SU(3)\otimes SU(2)_{L}\otimes SU(2)_{R}\otimes U(1)_{B-L}$ gauge symmetry. Second, the flipped variant of $SU(5)\otimes U(1)\subset SO(10)$ is the only one for which a radiative generation of the RH neutrino masses makes sense because in the standard $SU(5)$ the RH neutrinos are gauge singlets and as such they receive an explicit singlet mass term.

Besides this, the flipped scenario has got other virtues: the proton decay estimates\footnote{For a nice discussion on how to use BNV observables to distinguish between the standard and the flipped $SU(5)$ see, e.g., \cite{Dorsner:2004xx}.} may be under better control than in the standard $SU(5)$ because the leading theoretical uncertainties in the GUT-scale calculation (namely, the few-percent ambiguities in the GUT-scale matching of the running gauge couplings due to the Planck-induced effects~\cite{Hill:1983xh,Shafi:1983gz,Calmet:2008df,Chakrabortty:2008zk}) are absent. Furthermore, the flipped scenario offers better perspectives for a solution of the doublet-triplet splitting problem (if desired; see, e.g.,~\cite{Antoniadis:1987dx}) and, unlike in the ``standard'' $SU(5)$, there is no monopole problem in the flipped case either.

On top of that, the proposed scenario is in a certain sense even simpler than  the standard approach to the minimal\footnote{Minimality here refers to models without extra matter fields; for an alternative approach including, for instance, extra singlet fermions see, e.g.,~\cite{Abel:1989hq}.} renormalizable flipped $SU(5)$ where the seesaw scale is associated to the VEV of an extra scalar representation transforming as a 50-dimensional four-index tensor under $SU(5)$ coupled to the fermionic $10\otimes 10$ bilinear (see, e.g.,~\cite{Das:2005eb}) ; indeed, such a large multiplet is not necessary in the flipped $SU(5)$ \`a la Witten; as we shall argue, the two models can even be distinguished from each other if rich-enough BNV physics is revealed at future facilities.
In particular, we observe several features in the typical ranges predicted for the $\Gamma(p\to \pi^{0}e^{+})$ and  $\Gamma(p\to \pi^{0}\mu^{+})$ partial widths [as well as for those related by the isospin symmetry such as $\Gamma(p\to \eta e^{+})$ etc.] that are trivially absent in the model with $50_{H}$ in the scalar sector. Remarkably enough, this makes it even possible to obtain rather detailed information about all kinematically allowed $d=6$ nucleon decay channels in large portions of the parameter space where the theory is stable and perturbative.

The work is organized as follows: In Sec.~\ref{sectheory}, after a short recapitulation of the salient features of the standard and flipped $SU(5)$ models and the generic predictions of the partial proton decay widths
 therein, we focus on the Witten's loop as a means to constrain the shape of the (single) unitary matrix governing the proton decay channels into neutral mesons in the flipped case. In Sec.~\ref{secanalysis} we perform a detailed analysis of the simplest scenario in which a set of interesting correlations among the different partial proton decay widths to neutral mesons are revealed with their strengths governed by the absolute size of Witten's diagram. In Sec.~\ref{sectrealistic}, we adopt this kind of analysis to the minimal potentially realistic scenario. Then we conclude.

\section{$SU(5)\otimes U(1)$ \`a la Witten}\label{sectheory}
Let us begin with the basics of the flipped $SU(5)$ scheme and a short account of the $d=6$ proton decay in the $SU(5)$-based unifications focusing, namely, on the minimal versions of the standard and flipped scenarios and the potential to discriminate experimentally among them if proton decay would be seen in the future.
\subsection{The flipped $SU(5)$ basics}
The quantum numbers of the matter multiplets in the $SU(5)\otimes U(1)_{X}$  extensions of the canonical $SU(5)$ framework are dictated (up to an overall normalization factor) by the requirement of the gauge anomaly cancellation:
\be\label{charges}
\overline{5}_{M}\equiv (\overline{5},-3)\;, \;\;10_{M}\equiv (10,+1)\;,\;\;1_{M}\equiv (1,+5)\;.
\ee
Besides the ``standard'' $SU(5)$  assignment there is a second ``flipped'' embedding of the standard model (SM) hypercharge into the corresponding algebra, namely,
\be\label{hypercharge}
Y=\tfrac{1}{5}(X-T_{24})\,,
\ee
where the $SU(5)$ generator $T_{24}$ is in this case understood to conform the SM normalization (i.e., $Y=T_{24}$ and $Q=T_{L}^{3}+Y$ in the standard case). This
swaps $u^{c}\leftrightarrow d^{c}$ and $\nu^{c}\leftrightarrow e^{c}$ with respect to the standard $SU(5)$ field identification and, hence, the RH neutrinos fall into $10_{M}$ rather than\footnote{Recall that in the standard $SU(5)$ $Q$, $u^{c}$ and $e^{c}$ are in $10_{M}$, $d^{c}$ and $L$ in $\overline{5}_{M}$ and $\nu^{c}$ in $1_{M}$.} $1_{M}$. This also means that a VEV of a scalar version of $(10,+1)$ (to be denoted by $10_{H}$) can spontaneously break the $SU(5)\otimes U(1)_{X}$ gauge symmetry down to the SM\footnote{This is the observation in the core of the ``missing partner'' doublet-triplet splitting mechanism  (mainly relevant to supersymmetry) that brought a lot of  interest to the flipped $SU(5)$ scenario in the 1980s~\cite{Antoniadis:1987dx}.}.

Besides that, the scheme benefits from several nice features not entertained by the ``standard'' $SU(5)$ scenario, namely: (i)
The Yukawa Lagrangian
\be\label{welcome2}
{\cal L}\ni Y_{10}10_{M}10_{M}5_{H}+Y_{\overline{5}}10_{M}\overline{5}_{M}5^{*}_{H}+Y_{1}\overline{5}_{M}1_{M}5_{H}+h.c.\,,
\ee
including the 5-dimensional scalar $5_{H}=(5,-2)$ hosting the SM Higgs doublet, yields $M_{d}=M_{d}^{T}$, $M_{e}$ arbitrary and, in particular,  $M_{\nu}^{D}=M_{u}^{T}$, none of which is in a flagrant conflict with the observed quark and lepton flavor structure as it is the case for $M_{d}=M_{e}^{T}$ in the ``standard'' $SU(5)$. (ii)
The gauge unification is in a better shape than in the ``standard'' $SU(5)$ case because only the two non-Abelian SM couplings  are required to unify (which, indeed, they do at around $10^{16}$ GeV, cf. Appendix~\ref{RGEs})\nobreakdash --- note that the SM hypercharge is a ``mixture'' of the $T_{24}$ and $X$ charges (\ref{hypercharge}) and, thus, the SM coupling $g'$  obeys a nontrivial matching condition including an unknown coupling $g_{X}$ associated to the extra $U(1)_{X}$ gauge sector.  Hence, there is no need to invoke the TeV-scale supersymmetry for the sake of the gauge unification here as in the ``standard'' $SU(5)$ case.
(iii) Remarkably, the issue with the out-of-control Planck-scale induced shifts of the effective gauge couplings (and thus induced large uncertainties in the $M_{G}$ determination~\cite{Hill:1983xh,Shafi:1983gz,Calmet:2008df,Chakrabortty:2008zk}) is absent at the leading order because there is no way to couple the $10_{H}$ as the carrier of the large-scale VEV to the pair of the gauge field tensors $F_{\mu\nu}$. Thus, the prospects of getting a reasonably good grip on the proton lifetime in the flipped $SU(5)$ are much better than in the ordinary $SU(5)$ model.

The main drawback of such a scenario is the fact that the simplest ``conservative'' mechanism for generating a Majorana mass term for the RH neutrinos at the tree level requires an extra 50-dimensional scalar field $50_{H}\equiv(50,-2)$
whose large VEV in the $10_{M} 10_{M} 50_{H}$ contraction picks just the desired components\footnote{as does $\vev{\overline{126}_{H}}$ coupled to $16_{M}16_{M}$ in $SO(10)$}.
Obviously, one pays a big price here (i.e., 100 real degrees of freedom which further reduce the effective Planck scale~\cite{Larsen:1995ax,Veneziano:2001ah,Calmet:2008df,Dvali:2007hz}) and there is not much insight into the neutrino mass generation that this may provide (as, e.g.,  there is no grip on the flavor structure). Hence, this approach is not optimal as it totally ignores the bounty of the recent high-precision neutrino data.

\subsection{Proton decay in the standard and flipped $SU(5)$}
Since the new dynamics associated to the rich extra gauge and scalar degrees of freedom of the flipped $SU(5)$ scenario takes place at a very high scale the most promising observables it can find its imprints in are those related to the perturbative baryon number violation, namely, proton decay.

To this end, the flipped version of the $SU(5)$ unification is in a better shape than its ``standard'' cousin as it provides a relatively good grip~\cite{Barr:1981qv,Nath:2006ut} on the partial proton decay widths to neutral mesons and {\em charged} leptons whereas there is usually very little one can say on general grounds about these in the standard $SU(5)$ where those are the charged meson plus antineutrino channels which are typically under better theoretical control. Needless to say, this is very welcome as the observability of the charged leptons in the large-volume liquid scintilator~\cite{Autiero:2007zj}/water-Cherenkow~\cite{Abe:2011ts}/liquid Argon~\cite{Akiri:2011dv} experiments boosts the expected signal to background ratio and, hence,  provides a way better sensitivity (by as much as an order-of-magnitude) in these channels than in those with the unobserved final-state antineutrino.

Let us just note that this has to do, namely, with the hypercharge of the heavy $d=6$ proton-decay-generating gauge colored triplets which under the SM transform as $(3,2,-\tfrac{5}{6})$ in the standard $SU(5)$ case and as $(3,2,+\tfrac{1}{6})$ in the flipped $SU(5)$, respectively\footnote{This is also reflected by the classical notation where the $SU(2)$ components of the former are called $X$ and $Y$ while for the latter these are usually denoted by $X'$ and $Y'$.}. As for the former, the relevant $d=6$ effective BNV operators are~\cite{Nath:2006ut} of the ${\cal O}^{I}\propto \overline{u^{c}}Q\overline{e^{c}}Q$ and ${\cal O}^{II}\propto\overline{u^{c}}Q\overline{d^{c}}L$  type while for the latter these are ${\cal O}^{III}\propto\overline{d^{c}}Q\overline{u^{c}}L$ and ${\cal O}^{IV}\propto\overline{d^{c}}Q\overline{\nu^{c}}Q$ where ``pairing'' is always between the first two and the last two fields therein. Hence, the neutral meson+charged lepton decays in the standard $SU(5)$ receive contributions from both ${\cal O}^{I}$ and ${\cal O}^{II}$ while it is only ${\cal O}^{III}$ that drives it in the flipped scenario\footnote{In fact, ${\cal O}^{IV}$ is almost always irrelevant as it yields a left-hand antineutrino in the final state with typically (in the classical seesaw picture) a very tiny projection on the light neutrino mass eigenstates.}. On the other hand, the situation is rather symmetric in the charged meson+neutrino channels which in both cases receive sizeable contributions from only one type of a contraction [${\cal O}^{II}$ in $SU(5)$ and ${\cal O}^{III}$ in its flipped version]. Let us also note that the predictivity for these channels is further boosted by the coherent summation over the (virtually unmeasurable) neutrino flavors; hence, the inclusive decay widths to specific charged mesons are typically driven by the elements of the Cabibbo-Kobayashi-Maskawa (CKM) matrix.
For instance, in a wide class of simple $SU(5)$ GUTs (namely, those in which the up-type quark mass matrix is symmetric) the $p$-decay widths to $\pi^{+}$ and $K^{+}$ can be written as
\begin{align}
\Gamma(p\to \pi^{+}\overline{\nu})=&F_{1}|(V_{CKM})_{11}|^{2}M\,,\\
\Gamma(p\to K^{+}\overline{\nu})=&\left[F_{2}|(V_{CKM})_{11}|^{2}+F_{3}|(V_{CKM})_{12}|^{2}\right]M\nn\,,
\end{align}
where $F_{1,2,3}$ are calculable numerical factors and $M$ is a universal dimensionful quantity governed by the parameters of the underlying  ``microscopic'' theory such as the GUT scale, the gauge couplings, etc.
This feature is yet more pronounced in the simple flipped scenarios (namely, those in which the down-type quark mass matrix is symmetric\footnote{This, in fact, is the prominent case when the flipped-$SU(5)$ proton decay is robust, i.e., cannot be rotated away, cf.~\cite{Dorsner:2004jj,Dorsner:2004xx,Nath:2006ut}; for a more recent account of the same in a flipped-$SU(5)$ scenario featuring extra matter fields see, e.g.,~\cite{Barr:2013gca}.}); there one even obtains a sharp prediction
\begin{align}
\label{gamma1}\Gamma(p\to K^+\overline{\nu})&=0
\end{align}
which is a clear smoking gun of the flipped $SU(5)$ unification. For more details an interested reader is deferred to the dedicated analysis~\cite{Dorsner:2004xx}.

Coming back to the neutral meson channels in the simplest flipped $SU(5)$ scenarios (i.e., assuming symmetry of the down quark mass matrix), the partial widths of our main interest may be written in the form
\begin{align}
\label{gamma3}\frac{\Gamma(p\to \pi^0 e_\alpha^+)}{\Gamma(p\to \pi^+\overline{\nu})}&=\frac{1}{2}|(V_{CKM})_{11}|^2|(\PMNS U_\nu)_{\alpha 1}|^2\,,\\
\label{gamma4}\frac{\Gamma(p\to \eta e_\alpha^+)}{\Gamma(p\to \pi^+\overline{\nu})}&=\frac{C_2}{C_1}|(V_{CKM})_{11}|^2|(\PMNS U_\nu)_{\alpha 1}|^2\,,\\
\label{gamma5}\frac{\Gamma(p\to K^0 e_\alpha^+)}{\Gamma(p\to \pi^+\overline{\nu})}&=\frac{C_3}{C_1}|(V_{CKM})_{12}|^2|(\PMNS U_\nu)_{\alpha 1}|^2\,,
\end{align}
where $V_{PMNS}$ stands for the Pontecorvo-Maki-Nakagawa-Sakata leptonic mixing matrix and $U_{\nu}$ is the unitary matrix diagonalizing the light neutrino masses\footnote{Let us anticipate that Eqs~(\ref{gamma3})-(\ref{gamma5}) are written in the basis in which the up-type quark mass matrix is diagonal and real; needless to say, the observables of our interest are all insensitive to such a choice. For more details see Appendix~\ref{apbasis}.}. Note that $\PMNS U_\nu=U_e^L$ is the LHS diagonalization matrix in the charged lepton sector [see Eq.~\eqref{defPMNS}]; we write it in such a ``baroque'' way because $V_{PMNS}$ is measurable and, as will become clear, $U_\nu$ is constrained in the model under consideration.  The absolute scale in Eqs. (\ref{gamma3})-(\ref{gamma5}) is set by
\begin{align}
\label{gamma2}\Gamma(p\to \pi^+\overline{\nu})&=C_1 \left(\frac{g_G}{M_{G}}\right)^4\,,
\end{align}
where $g_{G}$ is the $SU(5)$ gauge coupling and the numerical factors
\begin{align}
C_1&=\frac{m_p}{8\pi f_\pi^2}A_L^2|\alpha|^2(1+D+F)^2\label{C1}\\
C_2&=\frac{(m_p^2-m_\eta^2)^2}{48\pi m_p^3 f_\pi^2}A_L^2|\alpha|^2(1+D-3F)^2\label{C2}\\
C_3&=\frac{(m_p^2-m_K^2)^2}{8\pi m_p^3 f_\pi^2}A_L^2|\alpha|^2\left[1+\frac{m_p}{m_B}(D-F)\right]^2\label{C3}
\end{align}
are obtained by chiral Lagrangian techniques, see~\cite{Nath:2006ut} (and references therein),~\cite{Dorsner:2004xx} and Appendix \ref{approton}.
From Eqs.~(\ref{gamma3})-(\ref{gamma5}), the theory's predictive power  for the proton decay to neutral mesons (especially for the ``golden'' $p\to \pi^{0}e^{+}$ channel), in particular, its tight correlation to neutrino physics, is obvious as the only unknown entry in Eqs. \eqref{gamma1}-\eqref{gamma5} is~the~unitary~matrix~$U_\nu$.

In what follows we shall exploit the extra constraints on the lepton sector flavor structure emerging in the flipped $SU(5)$ model with  Witten's loop  in order to obtain constraints on the admissible shapes of the $U_{\nu}$ matrix and, hence, get a grip on the complete set of proton decay observables. Let us note that this is impossible in the models in which the RH neutrino masses are generated in the ``standard'' way (e.g., by means of an extra $50_H$) where, due to the entirely new type of a contraction entering the lepton sector Lagrangian, $U_\nu$ typically remains unconstrained.

\subsection{Witten's mechanism in the flipped SU(5)}
The main benefit of dealing with a unification which is not ``grand'' (i.e., not based on  a simple gauge group) is the absence of the strict limits on the large-scale symmetry breaking VEVs from an overall gauge coupling convergence at around $10^{16}$~GeV. Indeed, unlike in the $SO(10)$ GUTs which typically require the rank-breaking VEV (e.g., that of 16- or 126-dimensional scalars) to be several orders of magnitude below $M_{G}$~\cite{Chang:1984qr,Deshpande:1992au,Deshpande:1992em,Bertolini:2009qj} and, hence, too low for Witten's loop to account for the ``natural''  $10^{12-14}$ GeV RH neutrino masses domain, no such issue is encountered in the $SU(5)\otimes U(1)$ scenario due to its less restrictive partial unification pattern. In particular, only the non-Abelian SM gauge couplings are supposed to converge toward $M_{G}$ which, in turn, should be compatible with the current proton lifetime limits; no other scale is needed. Furthermore, the $SU(5)\otimes U(1)$-breaking VEV $V_{G}\equiv \langle 10_{H}\rangle$ is perfectly fit from the point of view of the gauge structure of Witten's type of a diagram in this scenario.
\subsubsection{Witten's loop structure}
\begin{figure}[t]
\parbox{5cm}{\includegraphics[width=5cm,height=3cm]{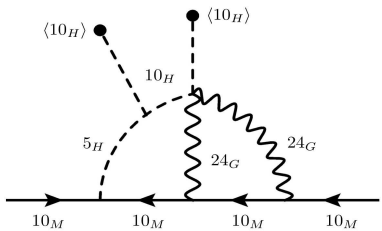}}
\caption{The gauge structure of Witten's loop in the flipped $SU(5)$ scenario under consideration. Note that we display just one representative out of several graphs that may be obtained from the one above by permutations.}
\label{figwitten}
\end{figure}
As in the original $SO(10)$ case the  gauge and loop structure of the relevant graphs (cf., Fig.~\ref{figwitten}) conforms\footnote{Note that the quantum numbers of the submultiplets under the $SU(5)$ subgroup of $SO(10)$ indicated in Witten's original work~\cite{Witten:1979nr} are irrelevant here as the RH neutrinos in the flipped scenario reside in $10$ of SU(5) rather than in a singlet.} several basic requirements: (i) there should be two $V_{G}$'s sticking out of the diagram so that the correct ``amount'' of the $U(1)_{X}$ breaking is provided for the desired RH neutrino Majorana mass term; (ii) the interactions experienced by the fermionic current must mimic the $10_{M}10_{M} 50_{H}$ coupling of the renormalizable  models in which the RH neutrino mass is generated at the tree level; (iii) only the minimal set of scalars required for the spontaneous symmetry breaking should be employed.
Given that, the structure depicted in Fig.~\ref{figwitten} turns out to be the simplest option\footnote{Note that minimality in this context depends on the specific construction of the perturbation expansion as, e.g., one diagram in the broken phase approach with massive propagators corresponds to a tower of graphs in the unbroken-phase theory when the VEVs are included in the interaction Hamiltonian.}; indeed, $5\otimes 24\otimes 24$ (where $24$ stands for the gauge fields) is the minimum way to devise the desired $50$. Note also that the $U(1)_{X}$ charge of the gauge $24_{G}$'s is zero and, thus, the two units of $X$ are delivered to the leptons via their Yukawa interaction with $5_{H}$. We have checked by explicit calculation that, indeed, the gauge structure of the graph yields a nonzero contribution for, and only for, the RH neutrino.
\subsubsection{The right-handed neutrino mass matrix}
Following the standard Feynman procedure the RH neutrino mass matrix can  be written in the form\footnote{Note that due to the symmetry of $Y_{10}$ the algebraic structure of the ``permuted'' graphs is the same as the one in Fig.~\ref{figwitten} and, hence, all contributions are covered by expression~(\ref{MnuMmain}).}
\be
M_{\nu}^{M}= \left(\frac{1}{16\pi^{2}}\right)^{2}g_{G}^{4}Y_{10}\,\mu\frac{\vev{10_{H}}^{2}}{M_{G}^{2}}\times {\cal O}(1)\label{MnuMmain}\,,
\ee
where $g_{G}$ is the (unified) gauge coupling corresponding to that of the $SU(5)$ part of the gauge group, $\mu$ is the (dimensionful) trilinear scalar coupling among $10_{H}$'s and $5_{H}$, cf.~Eq.~(\ref{potential}), $Y_{10}$ is the Yukawa coupling of $5_{H}$ to the matter bilinear $10_{M}\otimes 10_{M}$, cf.~Eq.~(\ref{welcome2}), $\vev{10_{H}}$ is the GUT-symmetry-breaking VEV, $M_{G}$ denotes the GUT scale and, finally, the ${\cal O}(1)$ factor stands for the remainder of the relevant expression. Besides the double loop-momentum integration (up to the geometrical suppression factors that have already been taken out in Eq.~(\ref{MnuMmain})) this may contain other structures specific for a particular evaluation method\footnote{Obviously, there are several equivalent approaches to the  evaluation of the momentum integrals involved in the ${\cal O}(1)$ factor: one can either work in the mass basis in which the propagators are diagonal and the couplings contain the rotations from the defining to the mass basis or vice versa; in principle, one may even work in a massless theory with VEVs in the interaction part of the Lagrangian.} such as, e.g., unitary transformations among the defining and the mass bases in different sectors. Note also that the second power of $M_{G}$ in the denominator is expected on dimensional grounds.

To proceed, we shall  cluster $g_{G}^{2}$ with the two powers of $V_{G}$ and formally cancel this against the $M_{G}^{2}$ factor (following the usual $M_{G}\sim g_{G} V_{G}$ rule of thumb)
\be
M_{\nu}^{M}= \left(\frac{1}{16\pi^{2}}\right)^{2}g_{G}^{2}Y_{10}\,\mu\, K\label{MnuMmain2}\,,
\ee
where the possible inaccuracy of this has been concealed into the definition of the (hitherto unknown) factor $K$.
This, in fact, is the best one can do until all the scalar potential couplings are fixed; since, however, we do not embark on a detailed analysis of the effective potential and its spectrum underpinning any possible detailed account for the relevant gauge unification constraints, all our results will be eventually parametrized by the value of $K$. A qualified guess of the size of the loop integral~\cite{Babuprivateconversation} (assuming no random cancellations) puts this factor to the ${\cal O}(10)$ ballpark; hence, in what follows we shall consider $K$ from about 5 to about 50.

In the rest of this section we shall elaborate on Eq.~(\ref{MnuMmain2}); although there are several undetermined factors there, namely, $Y_{10}$, $\mu$ and $K$, the former two are subject to perturbative consistency constraints following from the requirements of the SM vacuum stability and general perturbativity which, together with the above-mentioned bounds on $K$, impose rather strict limits on the absolute scale of the RH neutrino masses.
\subsubsection{Constraints from the SM vacuum stability}
Here we attempt to identify the parameter-space domains that may support a stable SM vacuum, i.e., those for which there are no tachyons (i.e., no negative-sign eigenvalues of the relevant scalar mass-squared matrix) in the spectrum.
\paragraph*{Tree-level scalar potential.}
Let us parametrize the tree-level scalar potential as
\bea
V_0&=&\frac{1}{2}m_{10}^{2}{\rm Tr}(10_{H}^{\dagger}10_{H})+m_{5}^{2}5_{H}^{\dagger}5_{H}\label{potential}\\
&+&\frac{1}{8}(\mu \varepsilon_{ijklm}10_{H}^{ij}10_{H}^{kl}5_{H}^{m}+h.c.)+\nn\\
&+&
\frac{1}{4}\lambda_{1}[{\rm Tr}(10_{H}^{\dagger}10_{H})]^{2}+
\frac{1}{4}\lambda_{2}{\rm Tr}(10_{H}^{\dagger}10_{H}10_{H}^{\dagger}10_{H})\nn\\&+&
\lambda_{3}(5_{H}^{\dagger}5_{H})^{2}+
\frac{1}{2}\lambda_{4}{\rm Tr}(10_{H}^{\dagger}10_{H})(5_{H}^{\dagger}5_{H})\nn\\&+&\lambda_{5}5_{H}^{\dagger}10_{H}10_{H}^{\dagger}5_{H}\nn\,,
\eea
where $10_{H}$ and $5_{H}$ are conveniently represented by a $5\times 5$ complex antisymmetric matrix and a $5$-component complex column vector, respectively, and the normalization factors in the interaction terms have been chosen such that they ensure simplicity of the resulting Feynman rules and, hence, of the results below. Note that we choose a basis in which the GUT-scale VEV $V_{G}$ and the electroweak VEV $v$ are accommodated in the following components:
\be\label{VEVs}
\vev{10^{45}}=-\vev{10^{54}}=V_{G}\,,\quad\vev{5^{4}}=v\,.
\ee

\paragraph*{The SM vacuum.}
The SM vacuum stationarity conditions read
\bea
V_{G}\left[m_{10}^{2}+V_{G}^{2}(2\lambda_{1}+\lambda_{2})+v^{2}(\lambda_{4}+\lambda_{5})\right]&=&0\,, \\
v\left[m_{5}^{2}+2v^{2}\lambda_{3}+V_{G}^{2}(\lambda_{4}+\lambda_{5})\right]&=&0\,.\nn
\eea
There are in general four solutions to this system, namely,
\bea
V_{G}=v=0 & : & SU(5)\otimes U(1)\,, \nn\\
V_{G}\neq 0, v=0 & : & SU(3)\otimes SU(2)\otimes U(1)\,,  \nn\\
V_{G}\neq 0, v\neq 0 & : & SU(3)\otimes U(1)\,,  \nn\\
V_{G}= 0, v\neq 0 & : & SU(4)\otimes U(1)\,,  \nn
\eea
with the preserved symmetry indicated on the right; the first three then correspond to consecutive steps in the physically relevant symmetry breaking chain.
\paragraph*{The scalar masses.}
As long as only the signs of the scalar mass-squares are at stakes one can work in any basis. Using the ``real field'' one, i.e., $\Psi=\{10^{*}_{ij},10^{ij},5^{*}_{i},5^{i}\}$, the mass matrix ${\cal M}^{2}\equiv \vev{\partial^{2}V/\partial \Psi^{*}\partial \Psi}$ evaluated in the SM vacuum has the following system of eigenvalues (neglecting all subleading terms):
\begin{align}
&m_{G_{1,\ldots,16}}^{2}=0\label{Goldstones}\\
&m_H^{2}=\left[4\lambda_{3}-\frac{2(\lambda_{4}+\lambda_{5})^{2}}{2\lambda_{1}+\lambda_{2}}\right]v^{2}\,, \label{Higgsmass}\\
&m_S^{2}=2(2\lambda_{1}+\lambda_{2})V_{G}^{2}\,,\label{Singlet}\\
&m_{\Delta_1}^{2}=-\tfrac{1}{2}(\lambda_{2}+\lambda_{5})V_{G}^{2}-\tfrac{1}{2}V_{G}\sqrt{(\lambda_{2}-\lambda_{5})^{2}V_{G}^{2}+4\mu^{2}},\nn\\
&m_{\Delta_2}^{2}=-\tfrac{1}{2}(\lambda_{2}+\lambda_{5})V_{G}^{2}+\tfrac{1}{2}V_{G}\sqrt{(\lambda_{2}-\lambda_{5})^{2}V_{G}^{2}+4\mu^{2}}\label{others}.
\end{align}
A few comments are worth making here:
\begin{itemize}
\item
The 16 zeroes in Eq.~(\ref{Goldstones}) correspond to the Goldstone bosons associated to the spontaneous breakdown of the $SU(5)\otimes U(1)$ symmetry to the $SU(3)_{c}\otimes U(1)_{Q}$ of the low-energy QCD$\otimes$QED,
\item
$m_H$ is the mass of the SM Higgs boson. Let us note that  the recent ATLAS~\cite{Aad:2012tfa} and CMS~\cite{Chatrchyan:2013yoa} measurements of $m_{H}$ indicate that the running effective quartic Higgs coupling at around $M_{G}$, i.e., the parenthesis in Eq.~(\ref{Higgsmass}), should be close to vanishing, see, e.g.,~\cite{Buttazzo:2013uya} and references therein,
\item
$m_S$ is the mass of the heavy singlet in $10_{H}$,
\item
The remaining eigenvalues correspond to the leftover mixture of the  colored triplets with the SM quantum numbers $(3,1,-\tfrac{1}{3})$ from $5_{H}\oplus 10_{H}$ (6 real fields corresponding to each eigenvalue).
\end{itemize}
\paragraph*{Absence of tachyons.}
Clearly, there are no tachyons in the scalar spectrum as long as
\bea
2\lambda_{1}+\lambda_{2}& >& 0\,,\\
2\lambda_{3}(2\lambda_{1}+\lambda_{2})& >& (\lambda_{4}+\lambda_{5})^{2}\,,\\
\lambda_{2}+\lambda_{5}& <& 0\,, \label{sum25}
\eea
and, in particular,
\be
|\lambda_{2}+\lambda_{5}|V_{G}> \sqrt{(\lambda_{2}-\lambda_{5})^{2}V_{G}^{2}+4\mu^{2}}\,,
\ee
which may be further simplified to
$
\mu^{2}< \lambda_{2}\lambda_{5}V_{G}^{2}.
$
Combining this with (\ref{sum25}) one further concludes that both $\lambda_{2}$ and $\lambda_{5}$ must be negative. This also means that $\lambda_{1}$ must be positive and obey $2\lambda_{1}>|\lambda_{2}|$ and, at the same time $\lambda_{3}$ must be positive. To conclude, the $\mu$ factor in formula~(\ref{MnuMmain2}) is subject to the constraint
\be
|\mu|\leq \sqrt{ \lambda_{2}\lambda_{5}}V_{G}\label{allimportant}
\ee
in all parts of the parameter space that can, at the tree level, support a (locally) stable SM vacuum.
\subsubsection{Perturbativity constraints}\label{sect:perturbativity}
Let us briefly discuss the extra constraints on the RHS of Eq.~(\ref{MnuMmain2}) implied by the requirement of perturbativity of the couplings therein. Since the graph in Fig.~\ref{figwitten} emerges at the GUT scale it is appropriate to interpret these couplings as the running parameters evaluated at $M_{G}$. Note that the effective theory below this threshold is the pure SM and, thus, one may use the known qualitative features of the renormalization group evolution of the SM couplings to assess their behavior over the whole domain from $v$ to $V_{G}$.

In general, one should assume that for all couplings perturbativity is not violated at $M_{G}$ and below $M_{G}$ the same holds for the ``leftover'' parameters of the effective theory. To that end, one should consider several terms in the perturbative expansion of all amplitudes in the relevant framework and make sure the (asymptotic) series thus obtained does not exhibit pathological growth of higher-order contributions (to a certain limit). This, in full generality, is clearly a horrendous task so we shall as usual adopt a very simplified approach. In particular, we shall make use of the fact that the  running of all dimensionless couplings in the SM is rather mild so, in the first approximation, it is justified to consider their values at only one scale and assume the running effects will not parametrically change them.  Hence, in what follows we shall assume that 
\be\label{perturbativityconstraints}
|\lambda_{i}|\leq 4\pi \quad \forall i
\ee for all the  couplings in the scalar potential.
\subsubsection{Resulting bounds on the $U_{\nu}$ matrix}
With this at hand one can finally derive the desired constraints on the $U_{\nu}$ matrix governing the proton decay channels to neutral mesons~(\ref{gamma3})-(\ref{gamma5}).
Indeed, using the seesaw formula, one can trade $M_{\nu}^{M}$ in Eq.~(\ref{MnuMmain2}) for the physical light neutrino mass matrix $m_{LL}$ and the Dirac part of the full $6\times 6$ seesaw matrix\footnote{Needless to say, there are always at least three RH neutrinos in the flipped $SU(5)$ models.} $
M_\nu^M=-M_\nu^{D} \left(m_{LL}\right)^{-1} (M_\nu^{D})^{T}$ which, due to the tight link between $M_\nu^{D}$ and the up-type quark mass matrix in the simplest scenarios,  $M_{\nu}^{D}=M_{u}^{T}$, yields $M_\nu^M=-M_u^{T} \left(m_{LL}\right)^{-1} M_{u}$. Furthermore,  the basis in the quark sector can always be chosen such that the up-quark mass matrix is real and diagonal (see Appendix~ \ref{apbasis}); at the same time, one can diagonalize $m_{LL}=U_\nu^T D_\nu U_\nu$ and obtain:

\be\label{MnuMseesaw}
M_\nu^M=-D_u U_\nu^\dagger D_\nu^{-1} U_\nu^* D_u\;.
\ee
Combining this with formula (\ref{MnuMmain2}) and implementing the vacuum stability constraint (\ref{allimportant}) one obtains 
\be\label{precentralformula}|D_u U_\nu^\dagger D_\nu^{-1} U_\nu^* D_u|\leq \frac{\alpha_{G}}{64\pi^{3}}\sqrt{\lambda_{2}\lambda_{5}}|Y_{10}|V_{G} K\,,
\ee
where we denoted $\alpha_{G}\equiv g_{G}^{2}/4\pi$.
Finally, assuming $\max_{i,j\in\{1,2,3\}}|(Y_{10})_{ij}|=1$ and saturating the perturbativity constraints  (\ref{perturbativityconstraints}) we have
\be\label{centralformula}
\max_{i,j\in\{1,2,3\}}|(D_u U_\nu^\dagger D_\nu^{-1} U_\nu^* D_u)_{ij}|\leq \frac{\alpha_{G}}{16\pi^{2}}V_{G} K\,,
\ee
which provides a very conservative global limit on the allowed form of $U_{\nu}$
and, hence, on the proton decay partial widths~(\ref{gamma3})-(\ref{gamma5}).
\subsubsection{Unification constraints}
Let us finish this preparatory section by discussing in brief the constraints from the requirement of the convergence of the running $SU(3)_{c}$ and $SU(2)_{L}$ gauge couplings at high energy which shall provide basic information about the scales involved, in particular, the approximate value of the $V_{G}$ parameter.  
Given (\ref{VEVs}), the $SU(2)_{L}$ doublet of the proton-decay-mediating colored triplet gauge fields $(X', Y')$ has mass $M_{G}=\tfrac{1}{2}g_{G}V_{G}$ while the mass of the heavy  $U(1)_{T_{24}}\otimes U(1)_{X}$ gauge boson (i.e., the one orthogonal to the surviving massless SM  $B$-field associated to hypercharge) reads $M_{B'}=2\sqrt{\tfrac{3}{5}g_{G}^{2}+g_{X}^{2}}V_{G}$ in the units in which the $U(1)_{X}$ generator is normalized as in Eqs.~(\ref{charges}) and~(\ref{hypercharge}).

Let us note again that in the flipped scenario of our interest the $M_{G}$ parameter corresponds to the scale at which the $(X', Y')$ are integrated into the theory in order to obey the $SU(3)_{c}$ and $SU(2)_{L}$ unification constraints. The specific location of this point and, thus, the absolute size of the proton decay width,  however, depends also on the position of the other thresholds due to the extra scalars to be integrated in at around $M_{G}$, in particular, the $SU(5)\otimes U(1)_{X}/SU(3)_{c}\otimes SU(2)_{L}\otimes U(1)_{Y}$ Goldstone bosons (\ref{Goldstones}), the heavy singlet (\ref{Singlet}) and the heavy colored triplets (\ref{others}). Rather than going into further details here we defer a dedicated analysis of the situation in Appendix~\ref{RGEs} and, in what follows, we shall stick to just a single reference scale  of $M_{G}=10^{16.5}$ GeV which corresponds to the lower limit obtained therein. This, in turn, yields $\Gamma^{-1}(p\to \pi^{+}\overline{\nu})$ of the order of $10^{38.5}$~years, cf. Fig.~\ref{fig:MG}. Remarkably enough, there is also an upper limit of the order of $10^{42}$ years which, however, is attained only in a ``fine-tuned'' region where the inequality (\ref{allimportant}) is saturated.

\section{A sample model analysis}\label{secanalysis}
In order to exploit formula (\ref{centralformula}), it is convenient to begin with its thorough inspection which, as we shall see, will provide a simple analytic information on the potentially interesting regions of the parameter space which will, subsequently, feed into the analysis of the BNV observables. Later on, we shall compare the analytics with results of a dedicated numerical analysis.

\subsection{Parameter space\label{sect:parameterspace}}
\paragraph*{1. CP conserving setup.}
For the sake of simplicity, we shall start with $U_\nu$ real orthogonal which  shall be parametrized by three CKM-like angles $\omega_{12}$, $\omega_{23}$ and $\omega_{13}$:
\bdm
U_\nu=U_{2\text{-}3}(\omega_{23})U_{1\text{-}3}(\omega_{13})U_{1\text{-}2}(\omega_{12})
\edm
where $U_{i\text{-}j}(\omega_{ij})$ stands for a rotation in the $i$-$j$ plane by an angle $\omega_{ij}$, e.g.
\be\label{Utheta}
U_{2\text{-}3}(\omega_{23})=\left(\begin{array}{ccc}
1&0&0\\
0&\cos{\omega_{23}}&\sin{\omega_{23}}\\
0&-\sin{\omega_{23}}&\cos{\omega_{23}}
\end{array}\right).
\ee
Assuming normal neutrino hierarchy we parametrize the (diagonal) neutrino mass matrix $D_\nu=\mathrm{diag}(m_1,m_2,m_3)$ by the (smallest) mass  $m_1$ of the mostly electronlike eigenstate. The other two masses are then computed from the oscillation parameters ($\Delta m_{A}^2=2.43\times 10^{-3}\,\mathrm{eV}^2$, $\Delta m_{\odot}^2=7.54\times 10^{-5}\,\mathrm{eV}^2$ \cite{Fogli:2012ua,Tortola:2012te}) and, for the sake of this study, we mostly ignore the uncertainties in these observables. Let us note that for the inverted hierarchy the analysis is technically similar but physically less interesting, see below.

As long as the ratios of $m_i^{-1}$'s are all below $m_t/m_c$, i.e., for $m_1\gtrsim 10^{-4}\,\mathrm{eV}$ (which we shall assume in the simple analysis below), the LHS of Eq.~\eqref{centralformula} is maximized for $\left(D_u U_{\nu}^\dagger D_{\nu}^{-1} U_{\nu}^* D_u\right)_{33} = m_t^2\left(U_{\nu}^\dagger D_{\nu}^{-1} U_{\nu}^*\right)_{33}$.
Hence, Eq.~(\ref{centralformula}) gets reduced to (using $V_{G}=2 M_{G}/g_{G}$)
\be\label{conel33}
\left(U_{\nu}^\dagger D_{\nu}^{-1} U_{\nu}^*\right)_{33} \leq K \frac{g_G}{32\pi^{3} m_t^2} \times 10^{16.5}{\rm GeV}\approx K\times 3 \, \mathrm{eV}^{-1}\,,
\ee
where we have taken\footnote{For further details see Appendix~\ref{RGEs}.} $g_{G}=0.5$.
Besides that, one gets
\be\label{el33}
\left(U_{\nu}^\dagger D_{\nu}^{-1} U_{\nu}^*\right)_{33}\! =\!
\frac{\sin^2\omega_{13}}{m_1} + \cos^2\omega_{13}\left(\!\frac{\sin^2\omega_{23}}{m_2}+\frac{\cos^2\omega_{23}}{m_3}\!\right),
\ee
which shows that the CKM-like parametrization of $U_{\nu}$ is very convenient because $\omega_{12}$ drops entirely from Eq.~\eqref{el33}.

For further insight, let us consider the extreme cases first. For $\omega_{13}=\omega_{23}=0$ (and for arbitrary $\omega_{12}$) one has
$
\left(U_{\nu}^\dagger D_{\nu}^{-1} U_{\nu}^*\right)_{33} = m_3^{-1},
$
whereas for $\omega_{13}=\omega_{23}=\frac{\pi}{2}$ the same equals to $m_1^{-1}$. While $m_3^{-1}$ ranges from $11\,\mathrm{eV}^{-1}$ to $20\,\mathrm{eV}^{-1}$ for all $m_1$'s lower than the current Planck and large-scale-structure limit of about\footnote{Note that this value corresponds to the Planck+BAO limit~\cite{Ade:2013zuv} quoted in~\cite{Riemer-Sorensen:2013jsa}, i.e., $\sum m_{\nu}<0.23$~eV at 95\% C.L.} $0.08\,\mathrm{eV}$~\cite{Riemer-Sorensen:2013jsa}, $m_1^{-1}$ may range in principle from $12\,\mathrm{eV}^{-1}$ to infinity. This explains why the latter setting may not be allowed by \eqref{conel33} if $m_1$ and $K$ are small enough.

\begin{figure}
\parbox{7cm}{\includegraphics[width=7cm]{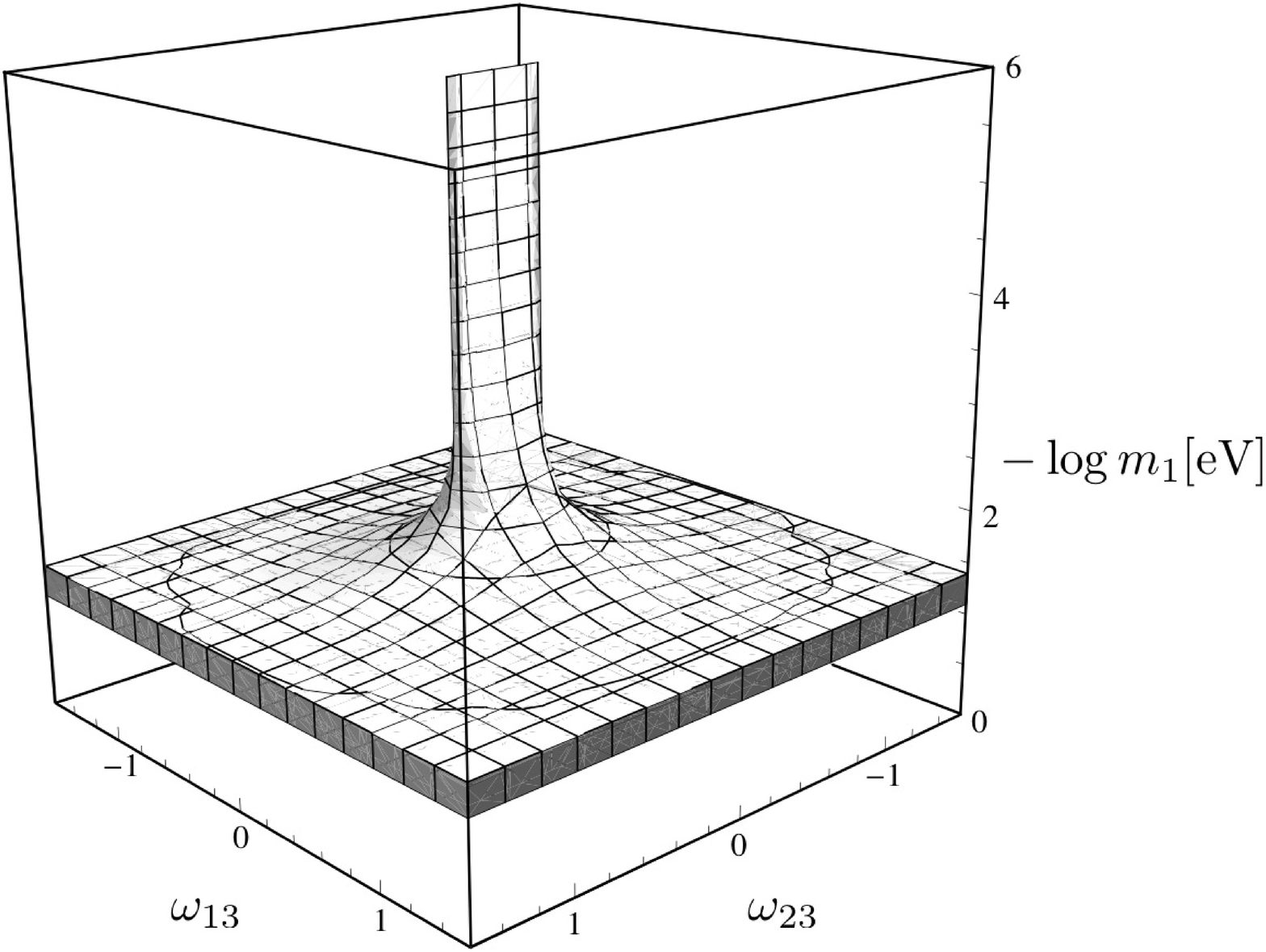}}\vskip 2mm
\parbox{7cm}{\includegraphics[width=7cm]{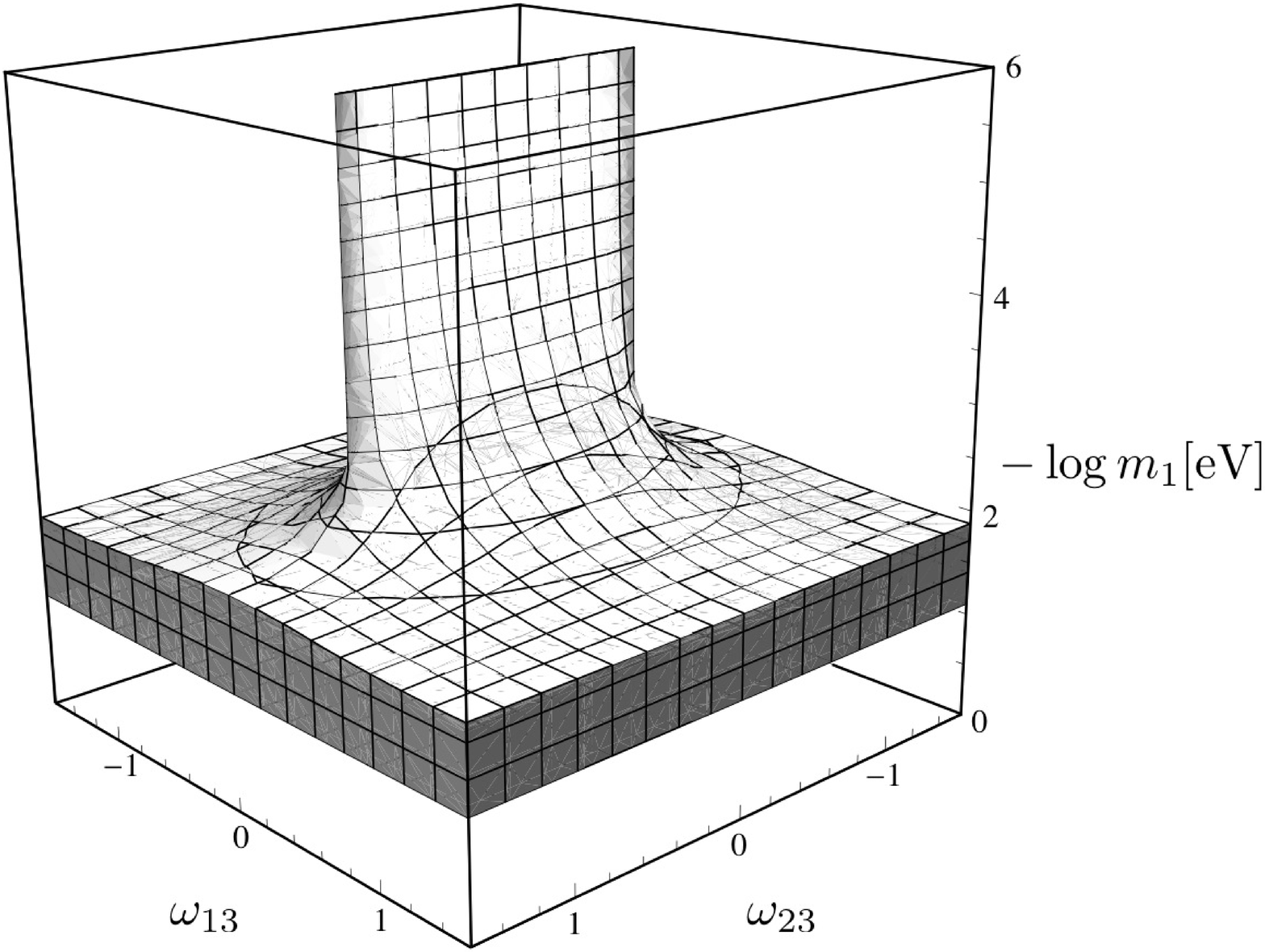}}
\caption{The shape of the allowed parameter space ($\omega_{23}$ and $\omega_{13}$ governing $U_{\nu}$ on the horizontal axes and the minus log of the lightest neutrino mass $m_{1}$ on the vertical; note that $m_{1}$ decreases from bottom to top) in the CP conserving setting discussed in Sec.~\ref{sect:parameterspace} for $K=10$ in the upper and $K=30$ in the lower panel, respectively. The allowed points are all those in the interior of the depicted structure. The straight cut in the lower part  corresponds to the current cosmology limit on the lightest neutrino mass $m_{1}\lesssim 8\times 10^{-2}$~eV~\cite{Riemer-Sorensen:2013jsa}, see the discussion in the text.}
\label{chimneysCPconserving}
\end{figure}

For the general case it is convenient to notice that the RHS of Eq.~\eqref{el33} is a convex combination of the inverse neutrino masses. Thus, for  $m_1^{-1}\leq K\times 3\, \mathrm{eV}^{-1}$ the inequality \eqref{conel33} is satisfied trivially. This can be clearly seen in Fig.~\ref{chimneysCPconserving} where the allowed parameter space is depicted: for $m_1\geq  (3 K)^{-1}\, \mathrm{eV}$, i.e, in the lower part of the plot, all $\omega_{23}$ and $\omega_{13}$ are are allowed. On the other hand, if $(m_3^{-1})_{\mathrm{min}} \approx 11 \, \mathrm{eV}^{-1} > K\times3\, \mathrm{eV}^{-1}$, i.e, if $K \lesssim 4$, \eqref{conel33} is never fulfilled.

There are two different regimes in the nontrivial region $m_1^{-1}\geq K\times3\, \mathrm{eV}^{-1} \geq m_3^{-1}$: if $m_1^{-1}\geq K\times 3\, \mathrm{eV}^{-1} \geq m_2^{-1}$ then for small enough $\omega_{13}$ any $\omega_{23}$ is allowed. More interestingly, for $m_2^{-1}\geq K\times 3\, \mathrm{eV}^{-1} \geq m_3^{-1}$, the allowed domain is confined to bounded regions around\footnote{Note that the RHS of Eq.~(\ref{el33}) is $\pi$-periodic.} $\omega_{13}=\omega_{23}=0$. This fully justifies the ``chimneylike'' shape in Fig. \ref{chimneysCPconserving} for $m_1^{-1}\geq K\times3\, \mathrm{eV}^{-1}$. It also follows that the allowed region becomes wider in the $\omega_{23}$ direction as $K$ grows, see again Fig.~\ref{chimneysCPconserving}. For $K$ above a certain critical value, the chimney would be wide open in the $\omega_{23}$ direction.

This is also why the results are less interesting for the inverted hierarchy -- there the two heavier neutrino masses are much closer to each other and, hence, the interesting region where $\omega_{13}$ and $\omega_{23}$ are constrained turns out to be too narrow.

\begin{figure}
\parbox{7cm}{\includegraphics[width=7cm]{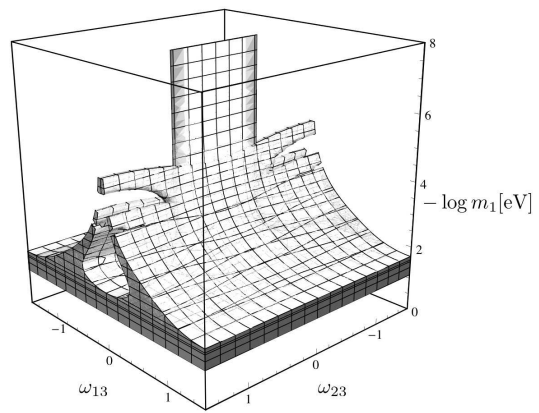}}
\caption{The same as in Fig.~\ref{chimneysCPconserving} for the CP violating setting with the ``Dirac'' phase in $U_{\nu}$ set to $\sigma=\pi/2$ and $K=20$. The net effect of a nonzero $\sigma$ is that $\omega_{23}$ remains unconstrained unless $m_{1}$ is really tiny [for which case the dominance of the 33 element in the RH neutrino mass formula~(\ref{MnuMseesaw}) is assumed].  The effects of the ``outer'' phases of $U_{\nu}$ in the observables discussed in Sec.~\ref{sect:parameterspace} are small so we conveniently fixed all of them to zero.}
\label{chimneysCPviolating}
\end{figure}

\paragraph*{2. CP violation.}
Second, let us discuss the case when $U_\nu$ is an arbitrary unitary matrix. In the CKM-like parametrization
\be\label{Uparametrization}
U_\nu=P_L U_{2\text{-}3}(\omega_{23}) U_{1\text{-}3}'(\omega_{13},\sigma) U_{1\text{-}2}(\omega_{12}) P_R\,,
\ee
where, as usual, $P_L=\mathrm{diag}\left(e^{i\rho_1}, e^{i\rho_2}, e^{i\rho_3}\right)$ and $P_R=\mathrm{diag}\left(1,e^{i\rho_4},e^{i\rho_5}\right)$ are pure phase matrices, $U_{2\text{-}3}(\omega_{23})$ and $U_{1\text{-}2}(\omega_{12})$ are as above, cf.~Eq.~\eqref{Utheta}, and $U_{1\text{-}3}'(\omega_{13},\sigma)$ contains an extra Dirac-like phase $\sigma$ analogous to the CP phase in the CKM matrix:
\bdm
U_{1\text{-}3}'(\omega_{13},\sigma)=\left(\begin{array}{ccc}
\cos{\omega_{13}}&0&\sin{\omega_{13}} e^{-i\sigma}\\
0&1&0\\
-\sin{\omega_{13}}e^{i\sigma}&0&\cos{\omega_{13}}
\end{array}\right).
\edm
It is clear that $\rho_4$ and $\rho_5$ drop from the $|(\PMNS U_\nu)_{\alpha 1}|$ combination in the decay rates \eqref{gamma3}-\eqref{gamma5} and, hence, they do not need to be considered. Since the analytics gets too complicated here let us just note that $\rho_1$, $\rho_2$ and $\rho_3$ play a very minor role in shaping the allowed parameter space and, thus, the only important phase in the game is $\sigma$; for $\sigma$ close to maximal the strict bounds on $\omega_{23}$ can be lost for much lighter $m_{1}$ than in the CP conserving case. As one can see in Fig. \ref{chimneysCPviolating}, for significant $\sigma$'s the $\omega_{23}$ parameter is typically out of control unless $m_1$ is taken to be very tiny [assuming again, for simplicity, the  dominance of the 33 element of the RH neutrino mass matrix (\ref{MnuMseesaw})].

\subsection{Observables}\label{sect:observables}
Since there is no $U_{\nu}$ in the partial proton decay widths to charged meson and the rates  \eqref{gamma4}-\eqref{gamma5} differ from \eqref{gamma3} only by calculable numerical factors let us focus here solely to  $\Gamma(p\to \pi^{0}\ell^{+})\equiv \Gamma_{\ell}$ for $\ell=e$, $\mu$.

It is not difficult to see that if $\omega_{23}$ can be arbitrary (such as in the lower parts of the allowed regions in Figs.~\ref{chimneysCPconserving} and \ref{chimneysCPviolating}) there is no control over $\Gamma_{\ell}$. However, if both $\omega_{13}$ and $\omega_{23}$ are restricted, there may be an upper bound on $|(\PMNS U_\nu)_{21}|$ and, hence, on $\Gamma_{\mu}$, while there is no such feature observed in $\Gamma_{e}$. On the other hand, there is a strong correlation among $\Gamma_{e}$ and $\Gamma_{\mu}$ which is clearly visible in the sum of the two decay rates; indeed, there is instead a lower bound on $\Gamma_{e}+\Gamma_{\mu}$. Hence, in what follows we shall stick to these two independent observables and note that very similar features can be seen in the decay rates to $K^0$  and $\eta$ related to these by the isospin symmetry.

To proceed, one also has to take into account that both $\Gamma_{\mu}$ and $\Gamma_{e}+\Gamma_{\mu}$  in general depend on $\omega_{12}$. Since, however, these relations are linear one can derive analytic expressions for  ``optimal'' $\omega_{12}$'s in each case such that $\Gamma_{\mu}$ is maximized and $\Gamma_{e}+\Gamma_{\mu}$ is minimized for any given values of  $\omega_{13}$ and $ \omega_{23}$.
Focusing, for simplicity, on the CP conserving case one has ($\PMNSx$ stands for the PMNS matrix)
\bdm
\tan{\omega_{12}^{\rm opt}}=\frac{\PMNSx_{23}\sin{\omega_{23}}-\PMNSx_{22}\cos{\omega_{23}}}{\PMNSx_{21}\cos{\omega_{13}}-\sin{\omega_{13}}\left(\PMNSx_{23}\cos{\omega_{23}}+\PMNSx_{22}\sin{\omega_{23}}\right)}
\edm
for the maximal value of $\Gamma_{\mu}$ (given $\omega_{13}$ and $ \omega_{23}$)
, whereas $\Gamma_{e}+\Gamma_{\mu}$ is (for given $\omega_{13}$ and $ \omega_{23}$) minimized for
\bdm
\tan{\omega_{12}^{\rm opt}}=\frac{\PMNSx_{33}\sin{\omega_{23}}-\PMNSx_{32}\cos{\omega_{23}}}{\PMNSx_{31}\cos{\omega_{13}}-\sin{\omega_{13}}\left(\PMNSx_{33}\cos{\omega_{23}}+\PMNSx_{32}\sin{\omega_{23}}\right)}\,.
\edm

\begin{figure}
\parbox{8.5cm}{\includegraphics[width=8.5cm]{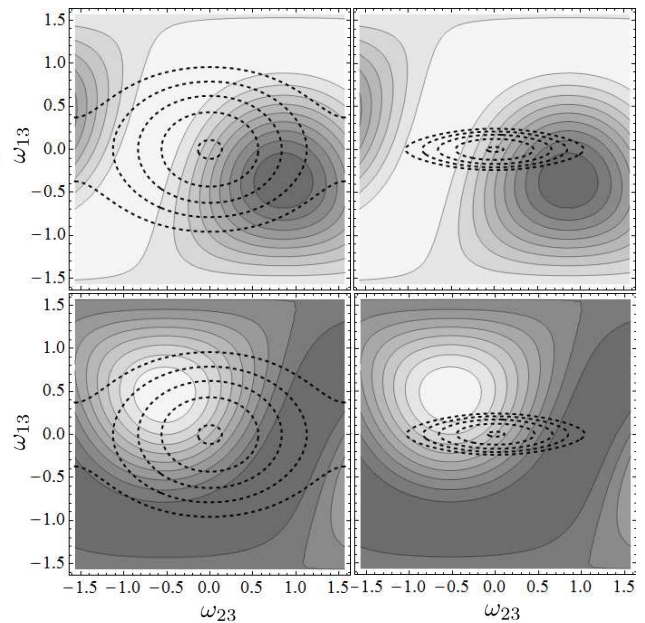}}
\caption{Contour plots of the $\omega_{12}$-extremes (cf. Sec.~\ref{sect:observables}) of the partial widths $\Gamma(p\to \pi^0 \mu^+)$ (upper panels, decreasing with darkening color) and $\Gamma(p\to \pi^0 e^+)+\Gamma(p\to \pi^0 \mu^+)$ (lower panels) superimposed with the (dashed) boundaries of the regions allowed by Eq.~\eqref{conel33} evaluated for $m_1=0.8\times10^{-2}\,\mathrm{eV}$ (left), and $m_1=0.8\times10^{-3}\,\mathrm{eV}$ (right), respectively. In all the plots the innermost and outermost dashed contours correspond to $K=7$ and $K=30$ respectively.}
\label{figcontours}
\end{figure}

In Fig. \ref{figcontours}, the solid contours in the upper two panels denote $\Gamma_{\mu}$ in units of $0.8 \times \tfrac{1}{2}\Gamma(p\to \pi^+\overline{\nu})|(V_{CKM})_{11}|^2\sim (10^{38} y)^{-1}$ (see Appendix~\ref{RGEs}) evaluated at the point $\{\omega_{12}^{\rm opt}(\omega_{23},\omega_{13}),\omega_{23},\omega_{13}\}$, i.e., at its upper limits for each $\omega_{23}$ and $\omega_{13}$; similarly, the lower limits on $\Gamma_{e}+\Gamma_{\mu}$ are displayed in the lower panels (the color code is such that the decay rates decrease in darker regions). At the same time, the dashed lines are boundaries of the regions allowed by \eqref{conel33} for different $K$'s, i.e., the ``horizontal cuts'' through different ``chimneys'' such as those in Fig.~\ref{chimneysCPconserving} at a constant $m_{1}$.

Remarkably enough, if $K$ is not overly large, there is a global upper limit on $\Gamma_{\mu}$, and a global lower limit on $\Gamma_{e}+\Gamma_{\mu}$ on the boundaries of the relevant allowed regions. Sticking to the $(-\pi/2,+\pi/2)$ interval for both $\omega_{13}$ and $\omega_{23}$, which is fully justified by the symmetry properties of the relevant formulas, the precise position of such a maximum (minimum) could be found numerically or well approximated by taking $\omega_{13}=0$ and the relevant $\omega_{23}$ on the boundary:
\be\label{theta23opt}
\cos^2{\omega_{23}}=\frac{m_2^{-1}- 3 K\,\mathrm{eV^{-1}}}{m_2^{-1}-m_3^{-1}}\,.
\ee
This formula holds for both observables, i.e., for the maximum of $\Gamma_{\mu}$ as well as for the minimum of $\Gamma_{e}+\Gamma_{\mu}$; one just has to choose $\omega_{23}\in(0,\pi/2)$ for the former and $\omega_{23}\in(-\pi/2,0)$ for the latter, respectively.

\subsection{Results}
\begin{figure}
\parbox{8.5cm}{\includegraphics[width=8.5cm]{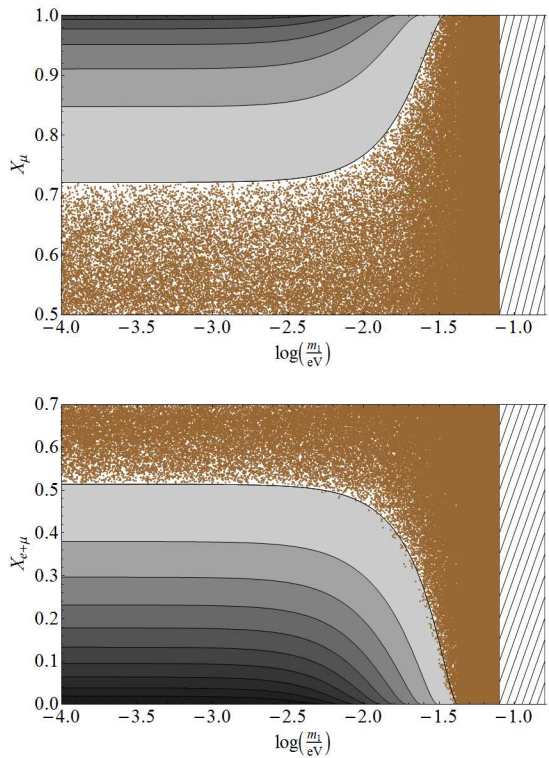}}
\caption{\label{fig:resultsmainCPconserved}The global upper limits on $X_{\mu}$ (upper plot) and the global lower limits on $X_{e+\mu}$ (lower plot), cf. Eqs.~(\ref{Xmu}) and (\ref{Xemu}),  as functions of the lightest neutrino mass (in the normal hierarchy case). The lowermost line on the upper plot, and the uppermost line on the lower plot correspond to $K=7$, with every consecutive contour for $K$ increased by $2$. The dots represent  an independent numerical calculation of the same decay rates for $K=7$ with  randomly chosen real $U_\nu$'s; only those points satisfying \eqref{centralformula} are permitted. The hatched area corresponds to $m_1>0.08\,\mathrm{eV}$ which is disfavored by cosmology~\cite{Riemer-Sorensen:2013jsa}.}
\label{figlinie}
\end{figure}

In what follows, we shall focus on a pair of observables $X_{\mu}$ and $X_{e+\mu}$ defined conveniently as
\bea\label{Xmu}
X_{\mu} & \equiv & \frac{\Gamma(p\to \pi^0 \mu^+)}{\tfrac{1}{2}\Gamma(p\to \pi^+\overline{\nu})|(V_{CKM})_{11}|^2}\,,\\
\label{Xemu}
X_{e+\mu} & \equiv & \frac{\Gamma(p\to \pi^0 e^+)+\Gamma(p\to \pi^0 \mu^+)}{\tfrac{1}{2}\Gamma(p\to \pi^+\overline{\nu})|(V_{CKM})_{11}|^2}\,;
\eea
their normalization (besides the trivial $|(V_{CKM})_{11}|^2$ piece) is fully governed  by the size of the $\Gamma(p\to \pi^+\overline{\nu})$ factor studied in detail in Appendix~\ref{RGEs}. 
\paragraph*{1. CP conserving case.}
If $U_\nu$ is real and orthogonal, both analytic and numerical analyses are tractable so it is interesting to see how these compare. In the upper plot in Fig. \ref{figlinie}, the solid lines indicate the analytic upper bounds on $X_{\mu}$ for a set of different $K$'s whereas the lower plots depict the corresponding lower bounds on $X_{e+\mu}$, respectively.

The points superimposed on both plots represent the results of a numerical analysis. For that sake, $m_1$ and the three CKM-like angles $\omega_{12}$, $\omega_{23}$ and $\omega_{13}$ were chosen randomly and we fixed $K=7$; only those points satisfying the inequality \eqref{centralformula} are allowed in the plot. We can see that, in spite of the simple $\omega_{13}=0$ assumption on the extremes of $X$'s, the analytic curves fit fairly well with the numerics. The agreement is slightly worse for larger $m_1$ which, however, is the case when the $\omega_{13}=0$ approximation becomes rather rough.\footnote{It is clear from Fig. \ref{figcontours} that the approximation of reaching the minimum at $\omega_{13}=0$ is more accurate for smaller $m_1$ (plots on the right-hand side) where the allowed regions are very narrow in the $\omega_{13}$ direction.}

\begin{figure}
\parbox{8.5cm}{\includegraphics[width=8.5cm]{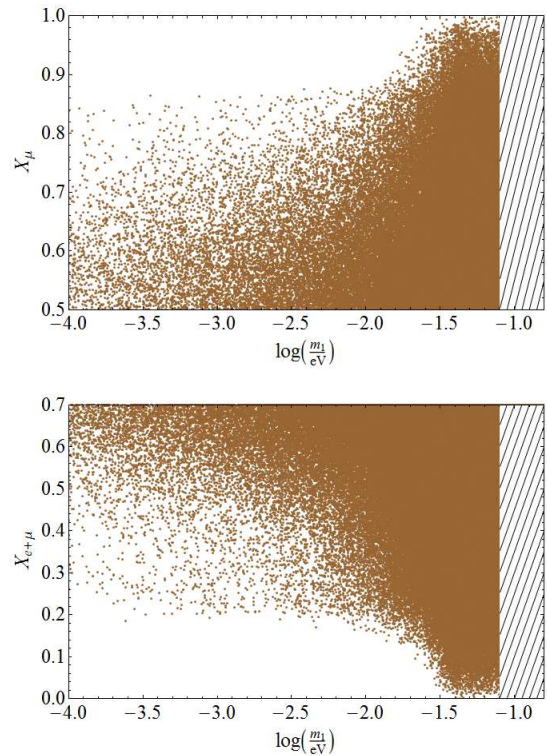}}
\caption{The same as in Fig.~\ref{fig:resultsmainCPconserved} but for a complex $U_{\nu}$ and $K=8$. The ``outer'' phases $\rho_1$, $\rho_2$ and $\rho_3$ (cf. Eq.~\ref{Uparametrization}) are varied randomly while the ``Dirac'' phase $\sigma$ of $U_{\nu}$ was fixed to zero. It is clear that the effect of $\rho_i$'s is very mild as the desired features in the partial widths remain essentially intact.}
\label{figbezdelta}
\end{figure}

\begin{figure}
\parbox{8.5cm}{\includegraphics[width=8.5cm]{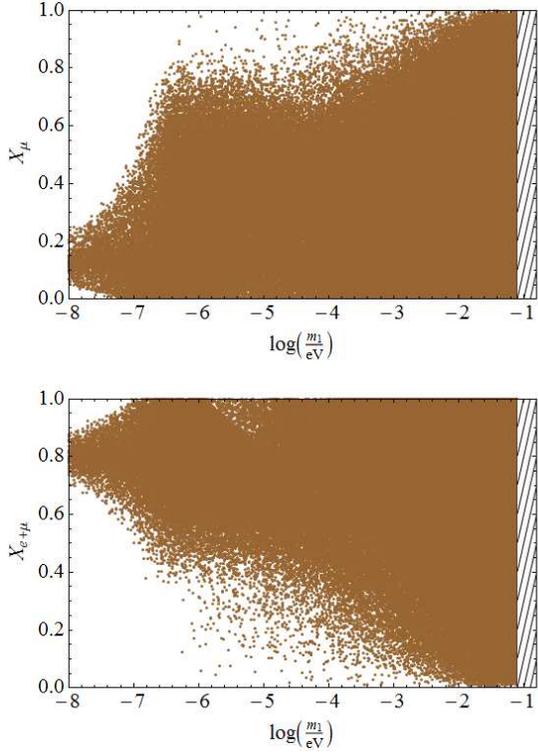}}
\caption{The same as in Fig. \ref{figbezdelta} but this time for entirely random phases in $U_{\nu}$ including $\sigma$. The effects in the partial widths are smeared until $m_1\lesssim 10^{-6}\, \mathrm{eV}$ because, for larger $m_{1}$, the important  constraints on $\omega_{23}$ from perturbativity and SM vacuum stability are lost, see Fig.~\ref{chimneysCPviolating}.}
\label{figvsechnyfaze}
\end{figure}

Concerning the physical interpretation of the results there are several options of how to read figure Fig. \ref{figlinie} and similar plots given in the next section. For instance, for a fixed $K$ (assuming, e.g., one can learn more about the high-scale structure of the theory from a detailed renormalization group analysis) a measurement of $X_{\mu}$ imposes a lower limit on mass of the lightest neutrino (e.g., $K=7$ and $X_{\mu} \sim 0.8$ is possible if and only if $m_{1}\gtrsim 10^{-2}$ eV etc.) Alternatively, for a given $K$ and a measured value of $m_{1}$ one gets a prediction for $X_{\mu}$ (for example, if $K=7$ and $m_{1}\sim 10^{-2}$ eV then $X_{\mu}$ is required to be below about $0.8$). Obviously, a similar reasoning can be applied to $X_{e+\mu}$.
\paragraph*{2. CP violation.}
The numerical analysis for a complex $U_{\nu}$ is far more involved and, besides that, there is no simple analytics that it can be easily compared to.  We allowed the three CKM-like angles and all the CP phases to vary arbitrarily within their domains and also $m_1$ was scanned randomly on the logarithmic scale. For  $\sigma$ close to zero one obtains similar features in $X_{\mu}$ and $X_{e+\mu}$ as in the CP conserving case regardless of the other three phases $\rho_1, \rho_2, \rho_3$, see Fig. \ref{figbezdelta}. If, however, also $\sigma$ is varied randomly, then both of these effects can be seen only for tiny  $m_1\lesssim 10^{-6}\, \mathrm{eV}$, cf. Fig.~\ref{figvsechnyfaze}. This, at least for the case of a dominant $33$ element of the RH neutrino mass formula~(\ref{MnuMseesaw}), can be easily understood from the shape of the allowed parameter space depicted on Fig. \ref{chimneysCPviolating}\nobreakdash ---there is no restriction on $\omega_{23}$ for moderate $m_{1}$ while for $m_1$ very tiny  $\omega_{13}$ and $\omega_{23}$ are again restricted to a bounded area.

\section{Potentially realistic scenarios}\label{sectrealistic}
A careful reader certainly noticed that, up to now, we have left aside the fact that in the most minimal model with only a single $5_{H}$ in the scalar sector the size of the Yukawa matrix entering  Witten's loop is further constrained by the need to reproduce the down-quark masses. Indeed, in such a case \be
Y_{10}\sim \tfrac{1}{\sqrt{2}}M_{d}/v\,,\label{smallY10}
\ee
which, barring renormalization group running, is at most of the order of $m_{b}/v\sim 2\%$. Hence, in the very minimal model  Witten's loop is further suppressed and the inequality~(\ref{centralformula}) cannot be satisfied unless $K$ is extremely large. In this respect, the perturbativity limits implemented in the discussion above are, strictly speaking, academic.

Another issue is the $M_{\nu}^{M}\propto M_{d}$ correlation which, regardless of the size of the proportionality factor, renders the light neutrino spectrum too hierarchical. Indeed, for $m_{LL}\propto M_{u}^{T} (M_{d})^{-1}M_{u}$
which in the $M_{d}$-diagonal basis  reads
$$
m_{LL}\propto W_{R}D_{u}V_{CKM}' (D_{d})^{-1}V_{CKM}'^{T}D_{u}W_{R}^{T}
$$
(provided $V_{CKM}'$ is the ``raw'' form of the CKM matrix including the five extra phases usually rotated away in the SM context and $W_{R}$ is an unknown unitary matrix) one typically gets $m_{2}:m_{3}\sim 0.001$ while the data suggest this ratio to be
close to $\sqrt{{\Delta m^{2}_{\odot}}/{\Delta m^{2}_{A}}}\sim  0.1$.
Hence, a potentially realistic generalization of the minimal scenario is necessary together with a careful analysis of the possible impacts of the extra multiplets it may contain on the results obtained in the previous sections.

There are clearly many options on how to avoid the unwanted suppression of $Y_{10}$ and get a realistic RH neutrino spectrum in more complicated models. One may, for example, add extra\footnote{There does not seem to be any loop-induced effect in the quark and/or charged lepton sectors of the original model that may provide the desired departure from the $M_{\nu}^{M}\propto M_{D}$ degeneracy; thus, extra degrees of freedom are necessary.} vectorlike fermions that may allow large $Y_{10}$ by breaking the correlation (\ref{smallY10}), heavy extra singlets etc.
However, in many cases the structure of such a generalized scheme changes so much that some of the vital ingredients of the previous analysis are lost.

In order to deal with this, let us first recapitulate the main assumptions behind the central formula (\ref{centralformula}) underpinning the emergence of all the features in the proton decay channels into neutral mesons seen in Sec.~\ref{secanalysis}:
First, the down-type quark mass matrix $M_{d}$ was required to be symmetric. This is not only crucial for the sharp prediction (\ref{gamma1}) but, on more general grounds, also to avoid the option of ``rotating away'' the $d=6$ gauge-driven proton decay from the flipped $SU(5)$ altogether, cf.~\cite{Dorsner:2004jj,Dorsner:2004xx,Nath:2006ut}. Second, in getting a grip on the LHS of Eq.~(\ref{MnuMmain}) we made use of the tight $M_{\nu}^{D}=M_{u}^{T}$ correlation. Obviously, both these assumptions are endangered in case one embarks on indiscriminate model building.

\subsection{The model with a pair of scalar $5$'s}\label{secpairof5s}
Remarkably enough, the simplest concievable generalization of all, i.e., the model with an extra $5$-dimensional scalar (which resembles the two-Higgs-doublet extension of the SM), renders the scheme perfectly realistic and, at the same time, it leaves all the key prerequisites of the analysis in Sec.~\ref{secanalysis} intact.

\subsubsection{The Yukawa sector and flavor structure}
Assuming both doublets in $5_{H}\oplus 5_{H}'$ do have nonzero projections onto the light SM Higgs the extended Yukawa Lagrangian
 \bea\label{welcome3}
{\cal L}&\ni &Y_{10}10_{M}10_{M}5_{H}+Y'_{10}10_{M}10_{M}5'_{H}+\nn\\
&+& Y_{\overline{5}}10_{M}\overline{5}_{M}{5}^{*}_{H}+Y'_{\overline{5}}10_{M}\overline{5}_{M}{5'_{H}}^{*}\nn\\
&+&Y_{1}\overline{5}_{M}1_{M}5_{H}+Y'_{1}\overline{5}_{M}1_{M}5_{H}'
+h.c.
\eea
gives rise to the following set of sum rules for the effective quark and lepton mass matrices
\bea
M_{\nu}^{D}&=&M_{u}^{T}\propto Y_{\overline{5}}v_{5}^{*}+Y'_{\overline{5}}v_{5'}^{*},  \\
M_{d}&=& M_{d}^{T}=Y_{10}v_{5}+Y'_{10}v_{5'}\label{Mdmain}\\
M_{e}&=& Y_{1}v_{5}+Y'_{1}v_{5'} \quad \text{arbitrary.}
\eea
Na\"\i vely, one would say that adding three extra 3$\times$3 Yukawa matrices (symmetric $Y_{10}'$, arbitrary $Y_{\overline{5}}'$ and $Y_{1}'$) the predictive power of the theory would be totally ruined.
However, from the perspective of the analysis in Secs.~\ref{sectheory} and~\ref{secanalysis} the only really important change is the presence of $Y_{10}'$; adding $Y_{\overline{5}}'$ and $Y_{1}'$ does not worsen the predictive power of the minimal setting at all because, for the former,  $M_{\nu}^{D}=M_{u}^{T}$ is still maintained and, for the latter, $M_{e}$ remains as theoretically unconstrained as before.

Indeed, the net effect of $Y_{10}'$ is just the breakdown of the unwanted $M_{\nu}^{M}\propto M_{d}$ correlation due to an extra term in the generalized version of formula (\ref{MnuMmain}):
\be
M_{\nu}^{M}= \left(\frac{1}{16\pi^{2}}\right)^{2}g_{G}^{4}(Y_{10}\,\mu+Y_{10}'\,\mu')\frac{\vev{10_{H}}^{2}}{M_{G}^{2}}\times {\cal O}(1)\label{MnuMmainGeneralized}\,.
\ee
Here $\mu'$ is the trilinear coupling of $5'_{H}$ to the pair of $10_{H}$'s analogous to the third term in formula~(\ref{potential}); as long as $\mu'/\mu$ is different enough from $v'/v$ one can fit all the down-quark masses without any need for a suppression in $Y_{10}$ and $Y_{10}'$.

Given this, the whole analysis in Sec.~\ref{secanalysis} can be repeated with the only difference that Eq.~(\ref{allimportant}) becomes more technically involved (but, conceptually, it remains the same) and, with that, there is essentially just an extra factor of 2 popping up on the RHS of the generalized formula (\ref{centralformula}):
\be\label{centralformulageneralized}
\max_{i,j\in\{1,2,3\}}|(D_u U_\nu^\dagger D_\nu^{-1} U_\nu^* D_u)_{ij}|\leq \frac{\alpha_{G}}{8\pi^{2}}V_{G} K
\ee
Hence, all results of Sec.~\ref{secanalysis} can be, in first approximation, adopted to the fully realistic case by a mere rescaling of the $K$ factor. For example, the allowed points depicted in Fig.~\ref{figbezdelta} for $K=8$ in the basic model are allowed in the generalized setting with $K=4$ and so on.

\section{Conclusions and outlook}

In this work we point out that the radiative mechanism for the RH neutrino mass generation, identified by E.~Witten in the early 1980s in the framework of the simplest $SO(10)$ grand unified models, can find its natural and potentially realistic incarnation in the realm of the flipped $SU(5)$ theory. This, among other things, makes it possible to resolve the long-lasting dichotomy between the gauge unification constraints and the position of the $B-L$ breaking scale governing  Witten's graph: on one side,  the current limits on the absolute light neutrino mass require $M_{B-L}$ to be close to the GUT scale which, on the other hand, is problematic to devise in the nonsupersymmetric unifications and even useless in the SUSY case where Witten's loop is typically canceled. In this respect, the relaxed unification constraints inherent to the flipped $SU(5)$ scheme allow not only for a natural and a very simple implementation of this old idea but, at the same time, for a rich enough GUT-scale phenomenology (such as perturbative baryon number violation, i.e.,  proton decay) so that the minimal model might be even testable at the near future facilities.

In particular, we have studied the minimal renormalizable flipped $SU(5)$ model focusing on the partial proton decay widths to neutral mesons that, in this framework, are all governed by a single unitary matrix $U_{\nu}$ to which one gets a grip through  Witten's loop. Needless to say, this is impossible in the usual case when the tree-level RH neutrino masses are generated by means of an extra 50-dimensional scalar and/or extra matter fields due to the general lack of constraints on the new couplings in such models. Hence, there are two benefits to this approach: the scalar sector of the theory does not require any multiplet larger than the 10-dimensional two-index antisymmetric tensor of $SU(5)$ and, at the same time, one obtains a rather detailed information about {\em all} $d=6$ proton decay channels in terms of a single and possibly calculable parameter.

To this end, we performed a detailed analysis of the correlations among the partial proton decay widths to $\pi^{0}$ and either $e^{+}$ or $\mu^{+}$ in the final state and we observed strong effects in the $\Gamma(p\to \pi^0 \mu^+)$ partial width (an upper bound) and in $\Gamma(p\to \pi^0 e^+)+\Gamma(p\to \pi^0 \mu^+)$ (a lower bound) across a significant portion of the  parameter space allowed by the perturbative consistency of the model, as long as normal neutrino hierarchy is assumed and the Dirac-type CP violation in the lepton sector is small. In other cases, such effects are observable only if the lightest neutrino mass is really tiny.

Concerning the strictness of the perturbativity and/or the SM vacuum stability constraints governing the size of these effects, there are several extra inputs that may, in principle, make these features yet more robust and even decisive for the future tests of the simplest models. If, for instance, proton decay would be found in the near future (at LBNE and/or Hyper-K) the implied upper limit on the unification scale (which, obviously, requires a dedicated higher-loop renormalization group analysis based on a detailed effective potential study) would further constrain the high-scale spectrum of the theory which, in turn, feeds into the computation of  Witten's loop and, thus, the $K$ factor; this, in reality, may be subject to stronger constraints than those discussed in Sec.~\ref{sectheory} with clear implications for the relevant partial widths. To this end, there are also other  high-energy signals that may be at least partially useful for this sake such as the baryon asymmetry of the Universe; although the $U_{\nu}$ matrix drops from the ``canonical'' leading order contribution to the CP asymmetry of the RH neutrino decays in leptogenesis, the size of the effective Yukawa couplings may still be constrained and, thus, also the $K$ factor.
This, however, is beyond the scope of the current study and will be elaborated on elsewhere.

\section*{ACKNOWLEDGMENTS}
The work of M.M. is supported by the Marie-Curie Career Integration Grant within the 7th European Community Framework Programme
FP7-PEOPLE-2011-CIG, Contract No. PCIG10-GA-2011-303565 and by the Research proposal MSM0021620859 of the Ministry of Education, Youth and Sports of the Czech Republic. The work of H.K. is supported by the Grant Agency of the Czech Technical University in Prague, Grant No. SGS13/217/OHK4/3T/14. The work of C.A.R. is in part supported by EU Network Grant No. UNILHC PITN-GA-2009-237920 and by the Spanish MICINN Grants No. FPA2011-22975, and No. MULTIDARK CSD2009-00064 and the Generalitat Valenciana (Prometeo/2009/091).~She is grateful for the hospitality of the Institute of Particle and Nuclear Physics of the Charles University in Prague during her visits in spring 2013.
We are indebted to Stefano Bertolini and Martin Hirsch for reading through the preliminary versions of the manuscript.

\appendix
\section{THE PROTON DECAY RATES}\label{approton}
In this appendix we rederive some of the results of paper \cite{Dorsner:2004xx} and rewrite them in our notation.
The proton decay partial widths to neutral mesons in the flipped $SU(5)$ model read
\begin{align}
&\Gamma(p\to \pi^0 e_\beta^+)=\frac{C_1}{2}\left|c(e_\beta,d^C)\right|^2\label{gammapi0}\,,\\
&\Gamma(p\to \eta e_\beta^+)=C_2\left|c(e_\beta,d^C)\right|^2\,,\label{gammaeta}\\
&\Gamma(p\to K^0 e_\beta^+)=C_3\left|c(e_\beta,s^C)\right|^2.\label{gammaK0}
\end{align}
with the constants $C_1, C_2, C_3$ defined in \eqref{C1}-\eqref{C3}. The $p$-decay widths to charged mesons obey
\begin{align}
&\Gamma(p\to \pi^+\overline{\nu})=C_1\label{gammapi+}\sum_{i=1}^3\left|c(\nu_i,d,d^C)\right|^2\,,\\
&\Gamma(p\to K^+\overline{\nu})=\sum_{i=1}^3\left|B_4 c(\nu_i,d,s^C)+B_5 c(\nu_i,s,d^C)\right|^2\label{gammaK+}\,,
\end{align}
where
\begin{align*}
B_4&=\frac{m_p^2-m_K^2}{2f_\pi\sqrt{2\pi m_p^3}}A_L |\alpha|\frac{2m_p}{3m_B}D\,,\\
B_5&=\frac{m_p^2-m_K^2}{2f_\pi\sqrt{2\pi m_p^3}}A_L |\alpha|\left[1+\frac{m_p}{3m_B}(D+3F)\right]\,,
\end{align*}
can be obtained from the chiral Lagrangian. The flavor structure of the basic contractions can be written like
\begin{align}
c(e_\alpha,d_\beta^C)&=k^2\left(U_d (U_u^L)^\dagger\right)_{\beta 1}(U_u^R (U_e^L)^\dagger)_{1\alpha},\label{ced}\\
c(\nu_l,d_\alpha,d_\beta^C)&=k^2(U_d U_d^\dagger)_{\beta\alpha}(U_u^R U_\nu^\dagger)_{1 l}.\label{cnudd}
\end{align}
Here $k={g_G}/{M_{G}}$ and the unitary matrices $U_d$, $U_u^{R,L}$, $U_\nu$ and $U_e^{R,L}$ provide the diagonalization of the quark and lepton mass matrices:
\begin{align*}
m_{LL}&=U_\nu^T D_\nu U_\nu\\
M_e&=(U_e^L)^T D_e U_e^R\\
M_d&=U_d^T D_d U_d\\
M_u&=(U_u^L)^T D_u U_u^R.
\end{align*}
Note that $M_d$ and $m_{LL}$ are symmetric, hence, instead of a biunitary, a  single-unitary-matrix transformation can be used to diagonalize each of them.
In this notation
\begin{align}
\label{defCKM}V_{CKM}&\propto U_u^L U_d^\dagger\\
\label{defPMNS}\PMNS&\propto U_e^L U_\nu^\dagger
\end{align}
where the proportionality sign turns into equality once the extra phases (unphysical from the SM perspective) are stripped down.
Hence, the flavor structure of the $d=6$ proton decay widths to neutral mesons and charged leptons is governed by
\be\label{coef}
\left|c(e_\alpha,d_\beta^C)\right|^2=k^4|(V_{CKM})_{1\beta}|^2|(U_u^R (U_e^L)^\dagger)_{1\alpha}|^2.
\ee

For a symmetric $M_{d}$ another important feature of the flipped $SU(5)$ scheme is recovered: $c(\nu_l,d_\alpha,d_\beta^C) \propto \delta_{\alpha\beta}$; this  implies $\Gamma(p\to K^+\overline{\nu})=0$. Moreover, considering $\sum_{l=1}^3 |(U_u^R U_\nu^\dagger)_{1 l}|^2=1$ one gets
\be
\Gamma(p\to \pi^+\overline{\nu})=\frac{m_p}{8\pi f_\pi^2}A_L^2|\alpha|^2(1+D+F)^2.
\ee

\section{THE CHOICE OF $M_{u}$-DIAGONAL BASIS}\label{apbasis}
It is convenient to choose the basis in which $M_{u}$ is diagonal, i.e., $U_u^L=U_u^R=\unit$. To justify this choice, we have to prove that all the quantities of our interest are independent of this choice. This concerns, in particular, the CKM and PMNS matrices and the proton decay widths \eqref{gammapi0}-\eqref{gammaK+}, i.e., the coefficient \eqref{coef}.

First, obviously, a transformation $U_u^L\rightarrow U_u^{L} V$ where $V$ is a unitary matrix must be compensated by a simultaneous change $U_d\rightarrow U_d V$ so that the CKM matrix \eqref{defCKM} remains intact.
Second, changing $U_u^R\rightarrow U_u^{R} W$ by a unitary $W$ requires $U_e^L\rightarrow U_e^L W$  otherwise \eqref{coef} is not preserved. On top of that, $U_u^R$ is related to $U_\nu$ via seesaw
$
m_{LL}=U_{\nu}^T D_{\nu} U_{\nu}=M_u^T\left(M_\nu^M\right)^{-1} M_u = -(U_u^R)^T D_u U_u^L\left(M_\nu^M\right)^{-1} (U_{u}^{L})^T D_u U_u^R,
$
hence also $U_\nu\rightarrow U_\nu W$ is induced. The transformations of $U_e^L$ and $U_\nu$ then act against each other so that also the PMNS matrix \eqref{defPMNS} remains unchanged. Thus, it is possible to choose $U_u^L=U_u^R=\unit$ without affecting any of the quantities discussed in Secs~\ref{sectheory} and~\ref{secanalysis}. In the $M_{u}$-diagonal basis the coefficient \eqref{coef} reads
\be\label{coef2}
\left|c(e_\alpha,d_\beta^C)\right|^2=k_2^4|(V_{CKM})_{1\beta}|^2|(\PMNS U_\nu)_{\alpha 1}|^2.
\ee
\section{$SU(3)_{c}\otimes SU(2)_{L}$ GAUGE UNIFICATION}\label{RGEs}
\begin{figure}[t]
\parbox{8.5cm}{\includegraphics[width=8.5cm]{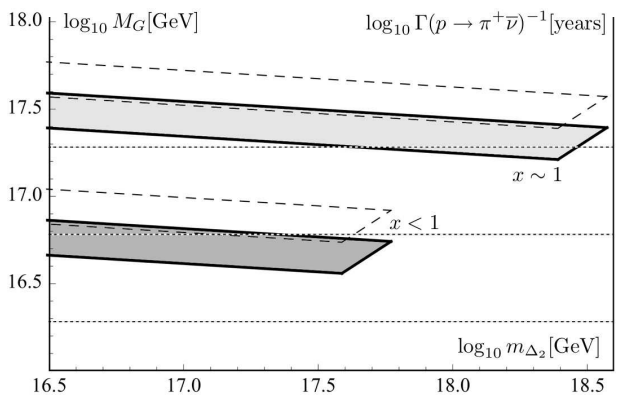}}
\caption{The unification constraints on the mass of the $(X',Y')$ gauge bosons (the left ordinate) and $\Gamma^{-1}(p^{+}\to \pi^{+}\overline{\nu})$ (the right ordinate) drawn for constant $x=\mu/\lambda V_{G}$ as functions of the masses of the scalar colored triplets $\Delta_{1}$ and $\Delta_{2}$, cf.~(\ref{others}) in the simplified case of $\lambda_{2}=\lambda_{5}$. The upper part of the plot corresponds to the ``fine-tuned'' region with $x$ very close to 1 with the mass of $\Delta_{1}$ significantly lower than $M_{G}$, cf. Eq.~(\ref{allimportant}), while the lower part corresponds to $x < 1$. The bands (one loop in dashed and two loops in solid) correspond to the 3-$\sigma$ uncertainty in $\alpha_{s}$ and their boundary on the right depicts the ``perturbativity'' limit $|\lambda_{i}|< 4\pi$, cf. Sec.~\ref{sect:perturbativity}.\label{fig:MG}}
\end{figure}
In order to get any quantitative grip on the absolute scale of the proton lifetime in the model(s) of interest, in particular, on $\Gamma(p^{+}\to \pi^{+}\overline{\nu})$ providing the overall normalization of the results depicted in Figs.~\ref{fig:resultsmainCPconserved}-\ref{figvsechnyfaze} one has to inspect thoroughly the constraints emerging from the requirement of the (partial) gauge coupling unification. Since the model is not ``grand'' unified in the sense that only the non-Abelian part of the SM gauge group is embedded into a simple component of the high-energy gauge group, this concerns only the convergence of the $g_{3}$ and $g$ couplings of the SM. 
Besides the ``initial condition'' defined by the values of $\alpha_{s}$ and $\alpha_{2}\equiv g^{2}/4\pi=\alpha/\sin^{2}\theta_{W}$ at the $M_{Z}$ scale and the relevant beta-functions the most important ingredient of such analysis is the heavy gauge and scalar spectrum shaping the evolution of $\alpha_{s}$ and $\alpha_{2}$ in the vicinity of $M_{G}$ [conveniently defined as the mass of the $(X',Y')$ gauge bosons] and, ultimately, their coalescence above the last of the heavy thresholds. 

As a reference setting let us start with the situation corresponding to the very simplest approximation in which all these heavy fields happen to live at a single scale ($M_{G}$); then, $M_{G}$ turns out to be at $10^{16.8}- 10^{17}$~GeV at one loop where the uncertainty corresponds to the 3-$\sigma$ band for $\alpha_{s}(M_{Z})$ and it gets reduced to about $10^{16.6}- 10^{16.8}$~GeV at two loops. 

Needless to say, such a single-mass-scale assumption is oversimplified as, in fact, the masses of the heavy colored triplet scalars $\Delta_{1}$ and $\Delta_{2}$, cf. Eq.~(\ref{others}) and the masses of the $(X',Y')$ gauge bosons [to quote only those states that are relevant here, i.e., $SU(3)_{c}\otimes SU(2)_{L}$ nonsinglets] depend on different sets of parameters and, hence, may differ considerably; this, in particular, applies for $\Delta_{1}$ that may be almost arbitrarily light if the inequality (\ref{allimportant}) gets saturated. This, obviously, may lead to a significant change in the ``na\"\i ve'' estimate above. 

In what follows, we shall focus on a simplified setting in which $\lambda_{2}=\lambda_{5}\equiv \lambda$  reflecting the symmetry of the relevant relations (\ref{others}) and (\ref{allimportant}) under their exchange and fix $g_{G}=0.5$. Hence, the masses of $\Delta_{1}$, $\Delta_{2}$ and $(X',Y')$ are fully fixed given $\lambda$, $\mu$ and $V_{G}$. This also means that if one fixes $m_{\Delta_{2}}$, $\lambda$ and $\mu$, then $m_{\Delta_{1}}$ and $M_{G}$ are fully determined and the unification condition can be tested. In turn, it can be used to get a correlation among the unification-compatible values of, say, $m_{\Delta_{2}}$ and $M_{G}$; the resulting situation is depicted in Fig.~\ref{fig:MG}. The shape of the allowed regions therein (in particular, the relatively shallow slope of the allowed bands for a fixed proportionality factor $x$ between $\mu$ and $\lambda V_{G}$) is easily understood: the effect of integrating in the $(X',Y')$ gauge bosons (plus the relevant Goldstones in the Feynman gauge) is much stronger than that of the two colored scalars $\Delta_{1,2}$ (assuming $x<1$, i.e., $m_{\Delta_{1}}$ not parametrically smaller than  $m_{\Delta_{2}}$) and, hence, a small shift in $M_{G}$ is enough to compensate even for significant changes in $m_{\Delta_{1,2}}$.

To conclude, the (two-loop) unification constraints limit the allowed domain of $M_{G}$ to the interval stretching from approximately $10^{16.5}$ GeV attained in the bulk of the parameter space up to about $10^{17.5}$ GeV if the fine-tuned configurations with $x\sim 1$ are considered. 


\end{document}